\documentclass[twocolumn, dvipsnames]{aastex631}
\usepackage{xcolor}
\usepackage{xspace}
\usepackage{float}

\usepackage{multirow}
\usepackage{hyperref}
\usepackage{subfigure}
\usepackage{amsmath}

\usepackage{graphicx}

\newcommand{\hmpc}{\,h^{-1}\,{\rm Mpc}}
\newcommand{\hmpcc}{\,h^{3}\,{\rm Mpc}^{-3}}
\newcommand{\hmsun}{\,h^{-1}\,{\rm M}_\odot}

\begin{document}

\title[Splashback galaxies and assembly bias]{The Impact of Splashback Galaxies on Galaxy Assembly Bias}
\author[0009-0001-1962-3924]{Hernan Rincon}
\affiliation{Department of Physics \& Astronomy, University of Rochester, 206 Bausch and Lomb Hall, P.O. Box 270171, Rochester, NY 14627-0171, USA}
\author[0000-0001-8286-6024]{Idit Zehavi}
\affiliation{Department of Physics, Case Western Reserve University, 10900 Euclid Avenue, Cleveland, OH 44106-1715, USA}
\author[0000-0001-7511-7025]{Sergio Contreras}
\affiliation{Facultad de F\'isica, Universidad de Sevilla, Campus de Reina Mercedes, Av. Reina Mercedes s/n 41012, Seville, Spain}
\author[0009-0002-6750-4919]{Yikun Wang}
\affiliation{Department of Physics, Case Western Reserve University, 10900 Euclid Avenue, Cleveland, OH 44106-1715, USA}
\author[0000-0002-9442-6808]{Xiaoju Xu}
\affiliation{Shanghai Key Lab for Astrophysics, Shanghai Normal University, Shanghai 200234, China}

\correspondingauthor{Hernan Rincon, Idit Zehavi}
\email{hrincon@ur.rochester.edu, idit.zehavi@case.edu}

\begin{abstract}
The clustering of galaxies is affected by the assembly history of their underlying dark matter halos. This complex phenomenon, known as galaxy assembly bias, has been extensively studied, but the exact physical origin remains unclear. Splashback halos, typically low-mass halos that have traversed larger neighboring halos, have been suggested to be associated with halo assembly bias. Using a semi-analytic galaxy-formation model applied to the Millennium simulation, we explicitly explore the role that splasback galaxies play in galaxy assembly bias. We identify splashbacks as present-day central galaxies that were formerly satellites of a more massive host, and construct stellar-mass selected galaxy samples with the splashbacks either removed or reclassified as satellites of their former host halo. 
We find that splashbacks tend to reside in low-mass, highly concentrated halos and in dense environments, and that they have relatively high stellar-to-halo mass ratios.  Splashbacks appear to be largely responsible for the low-mass tail of the occupancy variation for highly concentrated halos and for halos in dense environments. Finally, when computing the impact of assembly bias on galaxy clustering,  we find that while removing the splashbacks significantly reduces the signal, reassigning them has little effect on its amplitude but shifts the transition scale. We repeat the analysis with the hydrodynamical simulation TNG300, confirming the robustness of our results. Our results provide insight into assembly bias and have potential implications for modeling the galaxy-halo connection.
\end{abstract}

\section{Introduction}
In the standard cosmological framework of hierarchical structure formation, galaxies reside inside dark matter halos (e.g.,  \citealt{White:1978, White:1999, Wechsler:2018}).  The formation and evolution of the halos is dominated by gravity, while the formation of galaxies and their relation to the underlying dark matter also depend on complex physical processes. A description of the galaxy-halo relation is essential for understanding the physics of galaxy formation and for constraining models of dark energy through observations of galaxy clustering.

A central assumption that served as the basis for the development of halo modeling is that the galaxy content of halos depends only on halo mass and is statistically independent of the large-scale environment. This assumption originates from the uncorrelated nature of random walks describing halo assembly in the standard excursion set formalism \citep{Bond:1991, White:1999, Lemson:1999}. However, this assumption has been challenged by the demonstration in numerical simulations that the clustering of halos of fixed mass depends on secondary halo properties such as age, concentration, spin and tidal anisotropy (e.g., \citealt{Sheth:2004, Gao:2005, Wechsler:2006, Gao:2007, Salcedo:2018, Ramakrishnan:2019, Sato-Polito:2019}), a complex effect termed 
halo assembly bias. 

If galaxy properties correlate with halo assembly, we expect the galaxy content of halos to also depend on the secondary halo properties and the large-scale environment, leading to distinct variations in the halo occupation functions \citep{Zhu:2006, Artale:2018, McEwen:2018, Zehavi:2018, Bose:2019}. The predictions for these so-called occupancy variations have shown, for example, that halos in denser environments more readily host galaxies at lower halo mass. 

The combined impact of halo assembly bias and the occupancy variations results in a change in the large-scale amplitude of the galaxy correlation function \citep{Croton:2007}. This imprint of assembly bias on the galaxy distribution is often referred to as galaxy assembly bias (hereafter GAB). It is commonly studied in simulations by comparing the galaxy correlation function to that of a shuffled galaxy sample, where galaxies are randomly reassigned to halos of the same mass, effectively removing assembly bias (e.g., \citealt{Croton:2007, Zu:2008, ChavesMontero:2016, Zehavi:2018, Zehavi:2019, Contreras:2019, Contreras:2021a, Contreras:2021b, Xu:2021b, Xu:2021, Hadzhiyska:2021, Hadzhiyska:2023, Wang:2025b}). Neglecting these effects can have important implications for the interpretation of galaxy clustering (e.g., \citealt{Zentner:2014, Wang:2025}). 

The extent to which galaxies are affected by the assembly history of their host halos is an actively debated topic. The observational evidence for GAB remains inconclusive, given the challenges of constraining it without direct measures of halo properties. Several studies have claimed GAB detections (e.g., \citealt{Wang:2013b, Hearin:2015, Obuljen:2020, Yuan:2021, Contreras:2023, Oyarzun:2024, Pearl:2024,Ortega-Martinez:2025, Liu:2026, Rodriguez:2026}). However, other studies have indicated that the impact of assembly bias is small \citep{Blanton:2007, Lin:2016, Zu:2016, Walsh:2019, Salcedo:2022, Rocher:2023, Yuan:2024,Shao:2025} and that some of the previous claims were impacted by systematics \citep{Campbell:2015, Busch:2017, Sin:2017, Zu:2017, Sunayama:2019}.

The physical origins of assembly bias remain unclear. Different explanations have been proposed, such as correlated modes that break the random walk assumption, statistics of peaks, and truncation of low-mass halo growth in dense environments \citep{Keselman:2007, Sandvik:2007, Zentner:2007, Dalal:2008, Hahn:2009, Wang:2007, Ludlow:2011, Lacerna:2011, Zhang:2014, Borzyszkowski:2017, Mansfield:2020, Montero-Dorta:2024, Smith:2024, Montero-Dorta:2025}.

One possible origin of assembly bias is the ``arrested development'' of low-mass halos in the vicinity of massive neighboring halos. ``Splashback'' halos, in particular, pass within the inner regions of a massive halo, but are located outside its virial radius at the current epoch. Such halos may have their mass accretion truncated as a result of the tidal effects of the larger halo. Splashback halos are believed to have a significant role in halo assembly bias in the low-mass regime \citep{Dalal:2008, Hahn:2009, Wang:2009, Sunayama:2016, Mansfield:2020, Tucci:2021}. These halos are also sometimes referred to as ``backsplash'' or ``fly-by'' halos, with subtle distinctions in the exact definition (e.g., \citealt{Sinha:2012}). Here we refer to the galaxies within splashback halos as splashback galaxies (or simply splashbacks). 

Studies of splashback galaxies generally find that they are strongly impacted by environment-related effects, such as tidal disruption and ram-pressure stripping in cluster environments (see, e.g., \citealt{Balogh:2000, Gill:2005, Pimbblet:2011, Wetzel:2014, Zjupa:2020,  Borrow:2023, Ferreras:2023, Ruiz:2023}). Splashback galaxies have recently been associated with large-scale galactic conformity \citep{Lacerna:2018,Lacerna:2022,Ayromlou:2023,Wang:2023,Palma:2025}, a phenomenon possibly related to assembly bias \citep{Hearin:2015, Paranjape:2015, Lacerna:2025}. Splashback galaxies have also been suggested to contribute to anisotropic quenching in clusters (also known as angular conformity; \citealt{Martin-Navarro:2021, Stephenson:2025}).

We set out to investigate the galaxy-halo relation of splashback galaxies and their impact on GAB using a semi-analytical model
run on the Millennium Simulation. Though their impact on halo assembly bias has been explored, to the best of our knowledge, this is the first study aimed at directly investigating their role in galaxy assembly bias. 
The outline of the paper is as follows. Section \ref{sec:sims} describes the simulation and the galaxy populations used in our analysis. In Section \ref{sec:methods}, we describe our procedure for identifying simulated splashback galaxies and outline the GAB-related measures we analyze. Section~\ref{sec:results} presents our results, and we conclude in Section~\ref{sec:conclude}. We show that our results are robust against the modeling of baryonic physics in Appendix~\ref{sec:appendix}.

\section{Numerical simulation}
\label{sec:sims}

For the main part of this paper, we use stellar-mass selected galaxies from the \citet{Guo:2011} semi-analytical model applied to the Millennium Simulation \citep{Springel:2005}.  We also utilize the associated catalog of halos and subhalos along with their assembly history. We present the numerical simulation in \S~\ref{sec:sims.mil} and the semi-analytical model in \S~\ref{sec:sims.sam}. To demonstrate the robustness of our results to the galaxy-formation model, we also use the IllustrisTNG hydrodynamical simulation \citep{Pillepich:2018a}, as described in Appendix~\ref{sec:appendix}. 

\subsection{The Millennium Simulation}
\label{sec:sims.mil}

The Millennium Simulation is an N-body dark matter simulation of large-scale structure in a $\Lambda$CDM universe \citep{Springel:2005}. The simulation contains $2160^3$ particles in a $(500 \hmpc)^3$ periodic box, with a particle mass of $8.61 \times 10^8 \hmsun$. The positions of the particles evolve via gravitational interactions, calculated with the {\tt TREE-PM} N-body code {\tt GADGET-2}. The values of the cosmological parameters assumed in the simulation are $\Omega_{M}$ = 0.25, $\Omega_{b}$ = 0.045, $\sigma_8$ = 0.9, $n_s$ = 1 and $h$ = 0.73. The simulation has 64 outputs (or snapshots) between $z = 127$ and $z = 0$. Halos are identified using an {\tt FOF} halo-finding algorithm with a linking length of 0.2 times the mean particle separation \citep{Davis:1985}. 
Subhalos are identified using the topological properties of the density field with the {\tt SUBFIND} algorithm \citep{Springel:2001}. The final halo and subhalo catalogs allow for the construction of merger trees, which identify the progenitor and descendant halos of any given halo at a given snapshot.

We study in the following how the galaxy-halo connection varies with halo environment and halo concentration. These properties are chosen to allow  comparison with previous works assessing GAB \citep{Zehavi:2018, Contreras:2019, Xu:2021, Hadzhiyska:2021, Yuan:2021}. The environment $\delta_{1.25}$ is defined as the local dark matter particle density smoothed using a Gaussian filter with a scale of $1.25\hmpc$, interpolated to the halo locations from a grid of $2\hmpc$ cells. \citet{Zehavi:2018} have shown that the trends seen for occupancy variations with respect to environment are robust against the choice of smoothing scale by comparing the scales $1.25$, $2.5$, $5$, and $10\hmpc$. For our analysis, we choose to use the smallest smoothing scale, which has been shown to best characterize the level of GAB \citep{Xu:2021}. The halo concentration, $c$, is defined here as the ratio $V_{\text{max}}/V_{\text{vir}}$, where $V_{\text{max}}$ is the maximum circular velocity of the halo, and $V_{\text{vir}}$ is the circular velocity of the halo at the virial radius, $r_{\text{vir}}$. This definition serves as an easily computed proxy for the formal definition of halo concentration, $c=r_{\text{vir}}/r_{\text{s}}$, where 
$r_{\text{s}}$ is the scale radius of the \citet{Navarro:1997} profile fit to the halo (e.g., \citealt{Bullock:2001}).

\subsection{The Semi-Analytical Model}
\label{sec:sims.sam}

The \citet{Guo:2011} model is a version of {\tt L-GALAXIES}, the semi-analytic code of the Munich group \citep{Springel:2001, DeLucia:2004, Croton:2006, Bertone:2007, Guo:2013a, Henriques:2020}. The model is calibrated by fitting to observational data, such as the stellar mass function and luminosity function at low redshift, as well as the relation between black hole mass and bulge mass. The outputs of the model are publicly available on the Millennium database webpage\footnote{\url{https://wwwmpa.mpa-garching.mpg.de/millennium/}}.

We focus here on three galaxy samples with different number densities, ranked by the stellar mass of the galaxies. The three number densities are $0.00316 \hmpcc$, $0.01 \hmpcc$, and $0.0316 \hmpcc$, corresponding to stellar-mass thresholds of $3.88\times10^{10} \hmsun$, $1.42\times10^{10} \hmsun$, and $0.185\times10^{10} \hmsun$, respectively. 
The simulated samples are approximately evenly spaced in logarithmic number density, following the choices made in \citet{Zehavi:2018} and \citet{Xu:2021}. These stellar-mass thresholds provide populations well above the resolution limit of the simulation. They probe any dependence of the results on the stellar-mass limit, and allow for comparisons with realistic data samples that are limited by survey resolution. 

We chose this particular semi-analytic model and dark matter simulation because of the number of related works done with this simulation, including studies of GAB \citep{Busch:2017, Zehavi:2018, Xu:2021} and the large number of available properties for the halos, subhalos and environment of the simulation. While our choice of simulation assumes an older cosmological model, \citet{Contreras:2021b} has shown that the GAB signal is largely cosmology independent. We do not anticipate our results and overall conclusions to depend on these specific choices, as is also demonstrated in Appendix~\ref{sec:appendix}. By using a hydrodynamical simulation, we show that our results are not dependent on the modeling scheme used for central and satellite galaxies in the \citet{Guo:2011} SAM.

\section{Methods}
\label{sec:methods}

\subsection{Identifying Splashback Galaxies}
\label{sec:methods.definition}

Splashback halos are typically low-mass halos that have temporarily passed through larger neighboring host halos. Specifically, we may identify splashback halos using their ``fly-by'' assembly histories, which consist of three stages. They are initially outside the larger dark matter halo, later traverse within the larger halo’s virial radius, at which point they are classified as a subhalo, and finally their trajectory carries them out of the host halo once more. Alternatively, one may consider the ``splashback radius''  of (typically large cluster-sized) halos, namely the radius which encloses the apocenters of the first orbit of all accreted mass \citep{Adhikari:2014, More:2015, Rana:2023, Giocoli:2024, Xu:2024}. Splashback halos are then those exterior to a large halo's virial radius, but interior to its splashback radius. \cite{Mansfield:2020} have shown that both the splashback radius approach and the fly-by definition are consistent in identifying splashback halos as a cause of halo assembly bias. We adopt the latter for our analysis here.

The fly-by definition offers a relatively straightforward way to identify splashback galaxies using the simulation merger trees. In this scenario, the galaxies are initially centrals in their own halos, they later become satellites of the larger host halo and, finally, upon leaving that halo they return to their status as central galaxies. Thus, a practical definition of splashback galaxies is present-day central galaxies that were previously satellites of a currently more massive halo. The condition that the former host halo is more massive is imposed to exclude pathological major-merger cases, where the halos' relative sizes, defined in terms of their virial radii, may alternate repeatedly during the merger, creating artificial halo-subhalo-halo patterns. A similar definition for splashback halos is used in \citet{Mansfield:2020}. The splashback fraction that results from this definition for our various stellar mass cuts is detailed in Section \ref{sec:results.pop}. 

In what follows, we study the impact of the splashback galaxies on the different manifestations of GAB. We do this first by removing the splashback galaxies from the galaxy sample.  We also consider a scenario where we keep the splashback galaxies but reassign them as satellites of their former host halos. This is done under the premise that splashbacks may in fact be mislabeled satellites of their former host halo, as suggested by \citet{Diemer:2021}. In instances where the splashback central galaxies have their own satellites at present, these satellites are reassigned to the same former host halo. In cases where the former host halo is also a splashback, we recursively reassign the galaxies to a non-splashback host halo. 

The reassignment procedure serves to further examine the impact of the splashback galaxies. Namely, it addresses if the GAB signal depends on treating splashbacks as central galaxies of low-mass halos or as satellites of their former, more massive hosts. It also aids in assessing if splashback-induced assembly bias is model dependent. Different halo-finding algorithms, such as {\tt FOF} \citep{Davis:1985} and {\tt ROCKSTAR} \citep{Behroozi:2013}, categorize splashback halos differently. This introduces a potential algorithm-dependence of the splashback-related assembly bias. Comparing the results for a galaxy sample where the splashbacks are regarded as central galaxies in their own halos with a sample where they are re-classified as satellites of their former halos can provide insight on the model dependence. As shown in Section \ref{sec:results.effect}, we find that there is no significant model dependence, suggesting that analyses of splashback assembly bias are agnostic to the chosen halo finder. 

\subsection{Galaxy Assembly Bias Measures}  

To study the effects of assembly bias on the galaxy-halo connection, we examine the halo occupation function and its dependence on secondary properties beyond halo mass. While standard applications have assumed the mean halo occupation to depend just on halo mass, assembly bias introduces variations in the galaxy content of halos as a function of properties like halo age or concentration or the environment of the halos. We explore this in \S~\ref{sec:results} for the full galaxy samples, the samples with the splashbacks removed, and the samples with the splashback galaxies assigned to their former host halos.

To gain further insight, we also examine the stellar mass-halo mass relation for the central galaxies in the semi-analytic model applied to the simulation. Such diagnostics have provided valuable information on the physical origin of the occupancy variations \citep{Artale:2018,Zehavi:2018,Zehavi:2019}. Specifically, the scatter in the stellar mass-halo mass relation exhibits trends with the secondary properties which lead to the 
occupancy variations in the halo occupation functions. For example, at fixed halo mass,
older halos or halos in dense environments tend to host more massive galaxies \citep{Zehavi:2018}. We examine these trends here in the context of splashback galaxies.

To measure the impact of assembly bias on galaxy clustering in the simulated data sets, we compare the correlation function of the original galaxy sample to that of a shuffled sample with no assembly bias, following the methodology of \citet{Croton:2007}. We randomly reassign each halo's central galaxy to another halo within the same 0.1 dex halo-mass bin. The satellite galaxies are moved together with their original central galaxy, preserving their distribution around it (and thus maintaining the same one-halo contribution to the correlation function). This shuffling procedure removes the dependence of the halo occupation function on all secondary properties other than halo mass, and thus effectively eliminates GAB. We verified that our results are insensitive to the specific binning in halo mass.

Using the shuffling methodology, we measure the strength of GAB for the original samples, for the samples with splashback removed, and for samples with the splashbacks reassigned as satellites. We note that the relabeling of splashbacks as satellites of their former halos, in fact, does not change the clustering of the sample, since all galaxies maintain their spatial positions. However, whether the splashbacks are regarded as central galaxies or satellites does modify the shuffling procedure. In the reassignment case, the splashbacks (now satellites) move together with the central galaxy of their more massive former host halo. This change can impact the clustering of the shuffled sample and hence may lead to a difference in the amplitude of the GAB signal. We examine this below in \S~\ref{sec:results.effect}.

\section{Results}
\label{sec:results}

\subsection{The Splashback Galaxy Population}
\label{sec:results.pop}

We identify the splashback galaxies in the simulated galaxy population as described in \S~\ref{sec:methods.definition}. We will examine their impact for each of our three stellar-mass threshold samples (see \S~\ref{sec:sims.sam}).  The basic sample characteristics, including the number of splashback galaxies, for each of our three samples are provided in Table~\ref{tab:my_label}. 
The fraction of central galaxies that are splashbacks increases with the number density of the sample, varying from $2.5\%$ to $\sim8\%$. As higher number densities correspond to lower stellar mass thresholds, we see that splashbacks are more prevalent at the low-mass end of the central galaxies stellar-mass distribution. The fraction of all galaxies that are splashbacks likewise increases with decreasing stellar mass thresholds, from 1.6\% to 4.2\%. 

\begin{table}
    \centering
    \begin{tabular}{c|c|c|c|c|c|c|c}
    \hline
    n & ${\rm M}_{*}^{\rm thres}$ &${\rm N}_{\rm tot}$ & ${\rm N}_{\rm cen}$ & ${\rm N}_{\rm sat}$ & ${\rm N}_{\rm sp}$ & $f_{\rm tot}^{\rm sp}$ & $f_{\rm cen}^{\rm sp}$ \\
    \hline \hline
        0.0316 & $0.185$ & 3951 & 2129 & 1822 & 167 & 4.2 & 7.8 \\
        0.01 & $1.42$ & 1251 & 745 & 506 & 33 & 2.6 & 4.4 \\
        0.00316 & $3.88$ & 395 & 263 & 132 & 6.5 & 1.6 & 2.5 \\
    \end{tabular}
    \caption{Galaxy population numbers for samples of different stellar mass thresholds. Columns include the number density of the sample, the stellar mass threshold, total number of galaxies, number of central galaxies, satellites, and splashbacks, and the fraction of splashbacks relative to the full population and the centrals population. Number densities are quoted in units of $\hmpcc$, stellar mass thresholds are quoted in units of $10^{10} \hmsun$, number of galaxies are quoted in units of thousands, and fraction of splashbacks are quoted as a percentage.}
    \label{tab:my_label}
\end{table}

\begin{figure}
\includegraphics[width=0.45\textwidth]{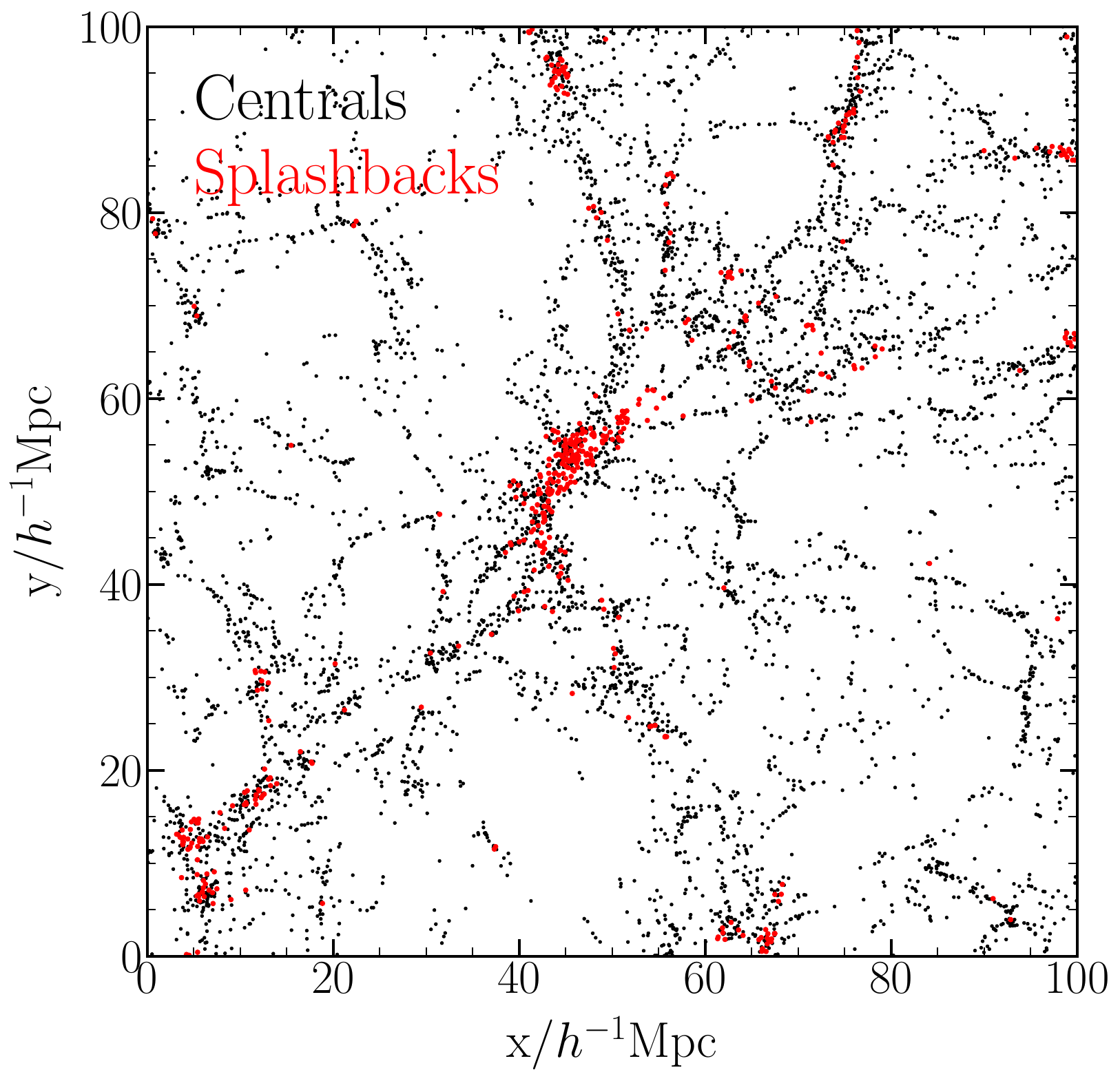} 
\includegraphics[width=0.45\textwidth]{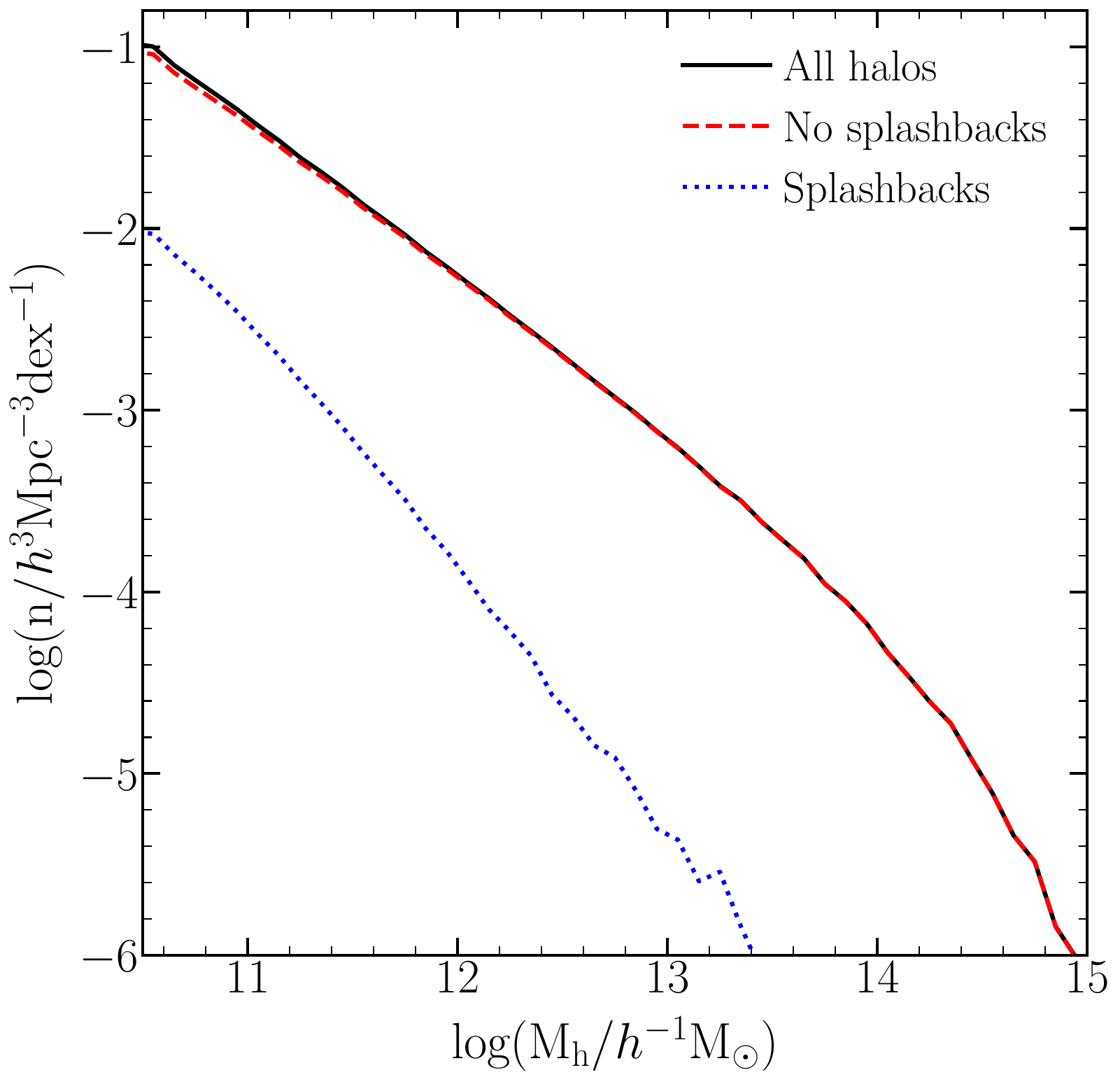}
\caption{The top panel shows the distribution of all central galaxies (halos) in a $5 \hmpc$ thick slice from the simulation. Splashback galaxies are denoted in red. It is apparent that they tend to reside in denser regions of the cosmic web. 
The bottom panel shows the halo mass function for all halos (solid black line), non-splashback halos (dashed red line) and 
splashback halos (dotted blue line). 
} 
\label{Fig:CW}
\end{figure}

An illustration of the spatial distribution of the splashback galaxies is shown in the top panel of Figure~\ref{Fig:CW}. We plot the position of all central galaxies (halos) in a $5 \hmpc$ thick slice of the simulation, with splashback galaxies denoted in red.   
Qualitatively, splashbacks appear to trace the dense cluster regions of the simulation's cosmic web. This preference for dense environments is expected, as splashbacks result from interactions with massive halos, which preferentially occur in denser regions. Given that galaxies in denser large-scale environments are more highly biased \citep{Bardeen1986}, we may then expect that splashback galaxies are highly biased objects.

The bottom panel of Figure~\ref{Fig:CW} shows the halo mass function of the Millennium Simulation. The black solid line includes all halos, while the red dashed line corresponds to non-splashback halos only, and the blue dotted line shows the splashback halos. It is apparent that splashbacks preferentially occupy the low-mass end of the halo mass distribution. This is expected given their ``fly-by" nature of traversing a more massive halo and the resulting arrested mass development. With the small fraction of splashback halos, the overall effect of excluding them (as shown by the red dashed line) is negligible relative to the full halo population.

\begin{figure*}
    \centering
    \subfigure[]{\includegraphics[width=0.45\textwidth, height=0.26\textheight]{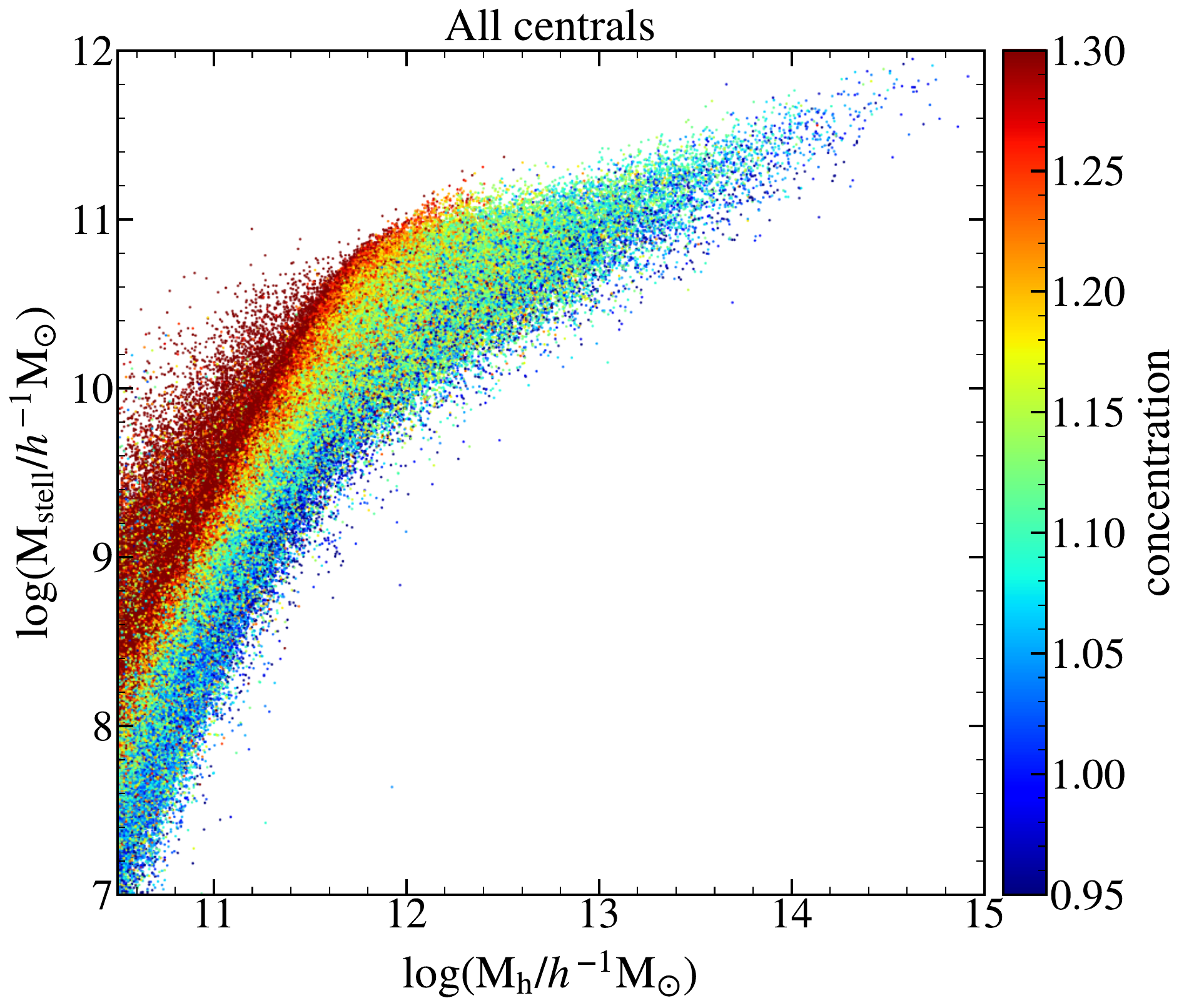}} 
    \subfigure[]{\includegraphics[width=0.45\textwidth, height=0.26\textheight]{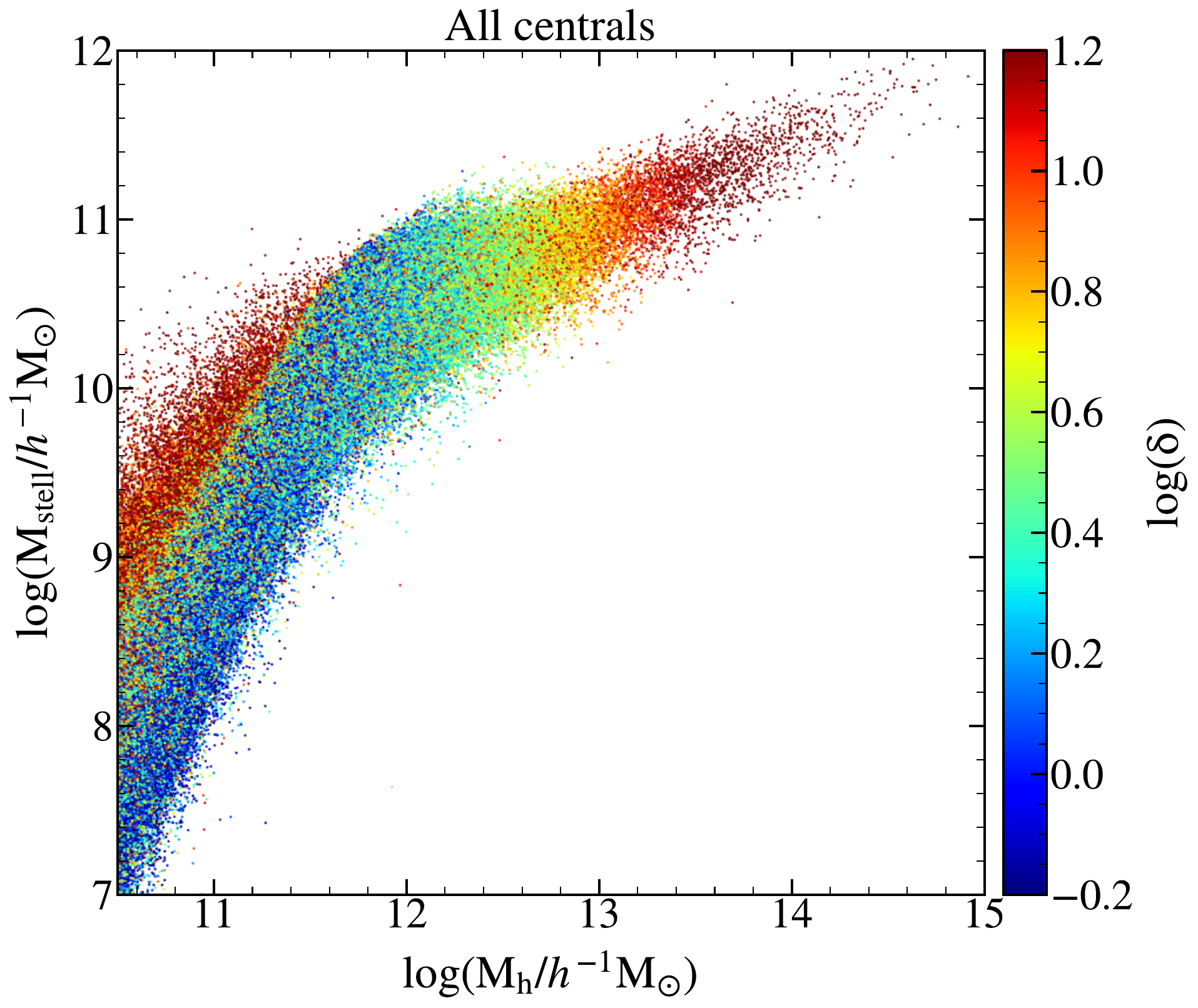}} 
    \subfigure[]{\includegraphics[width=0.45\textwidth, height=0.26\textheight]{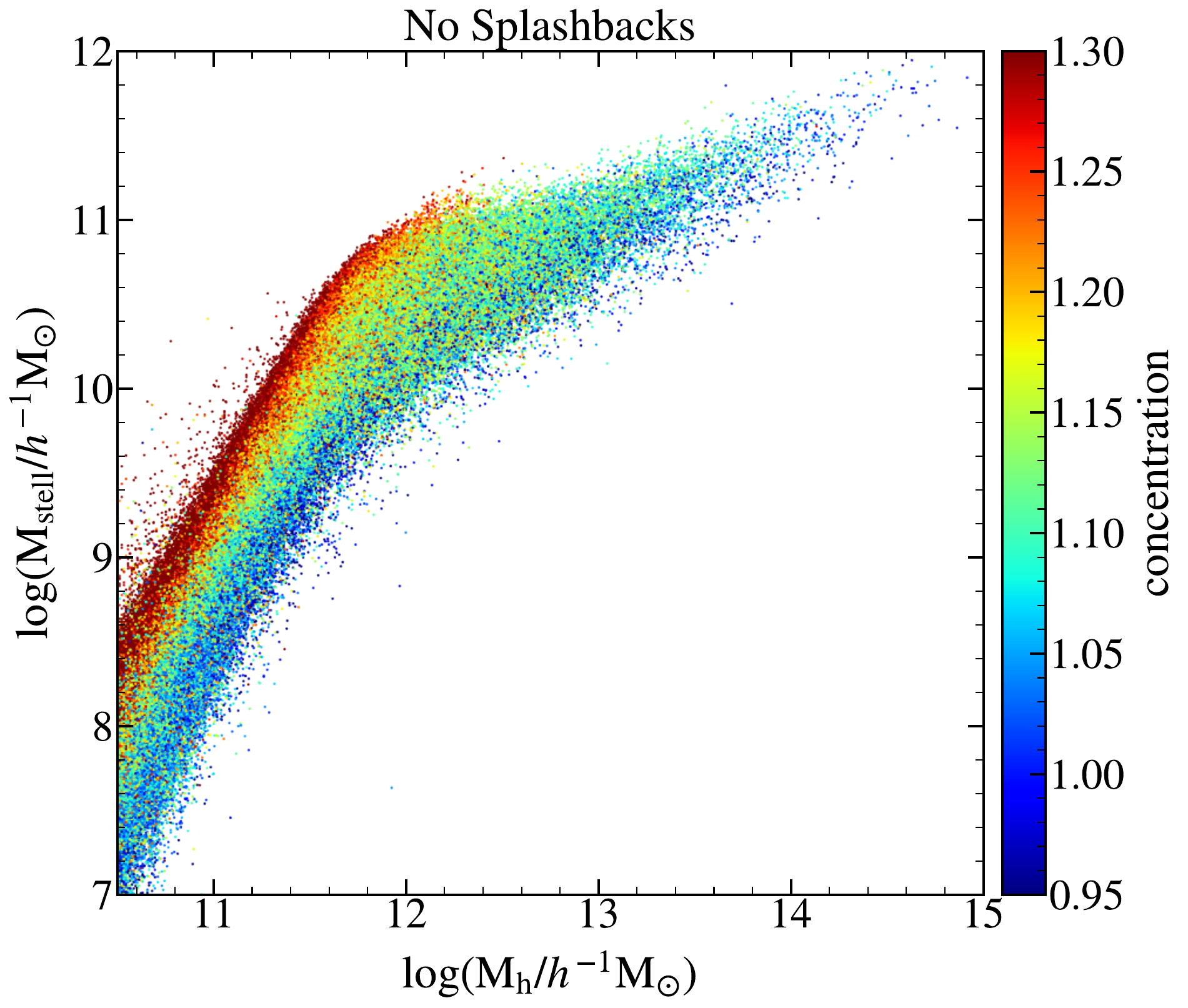}}
    \subfigure[]{\includegraphics[width=0.45\textwidth, height=0.26\textheight]{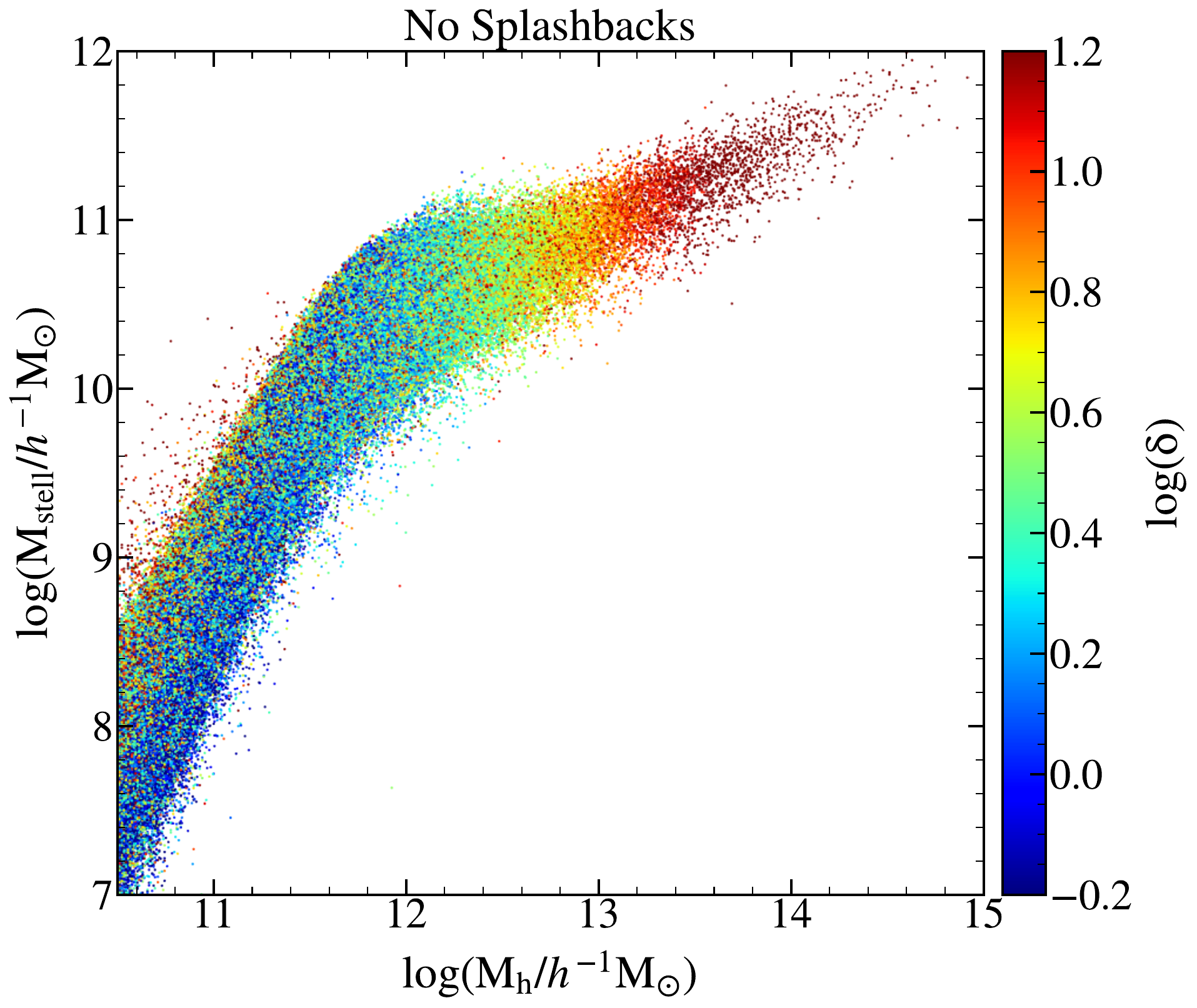}}
    \subfigure[]{\includegraphics[width=0.45\textwidth, height=0.26\textheight]{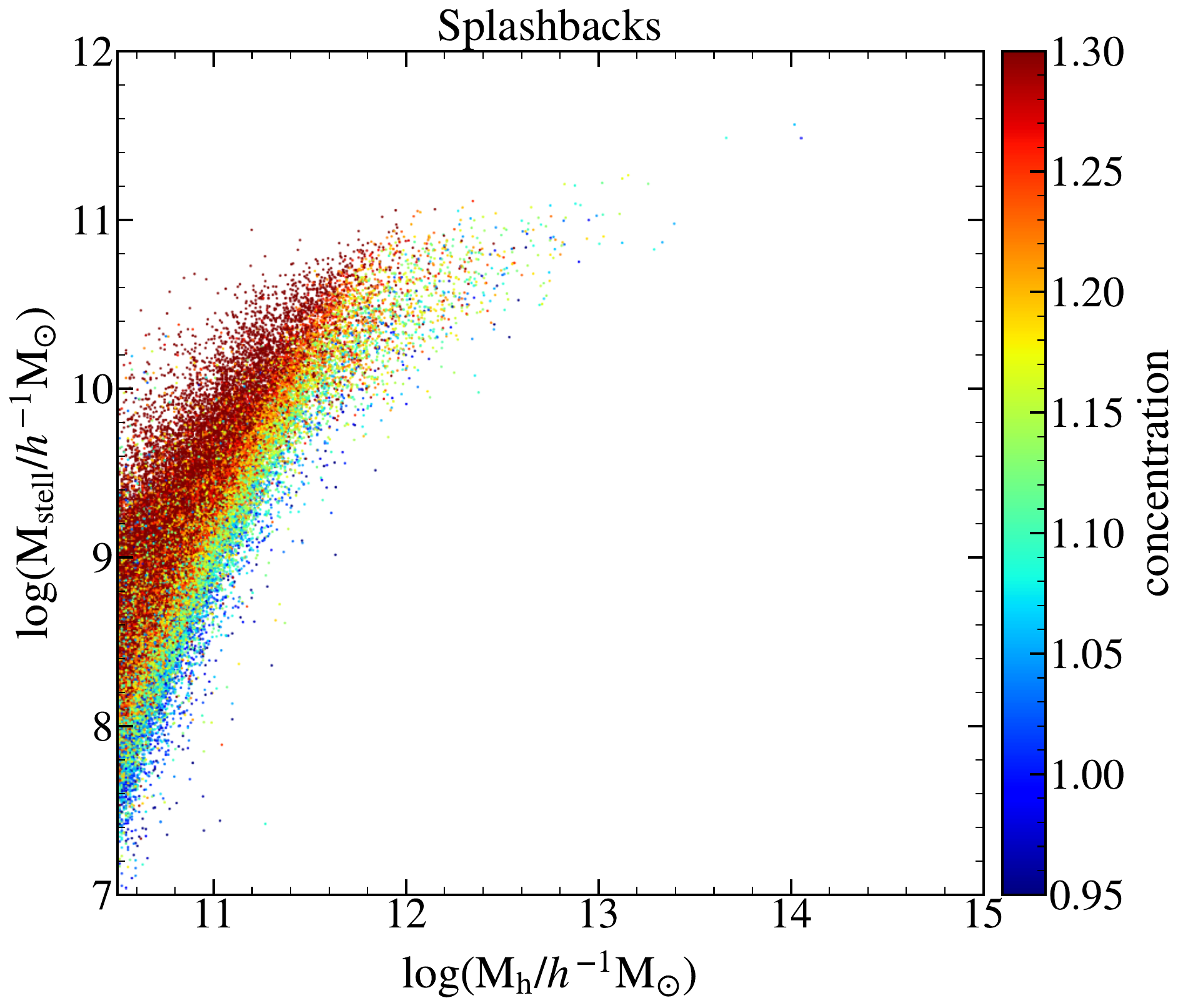}}
    \subfigure[]{\includegraphics[width=0.45\textwidth, height=0.26\textheight]{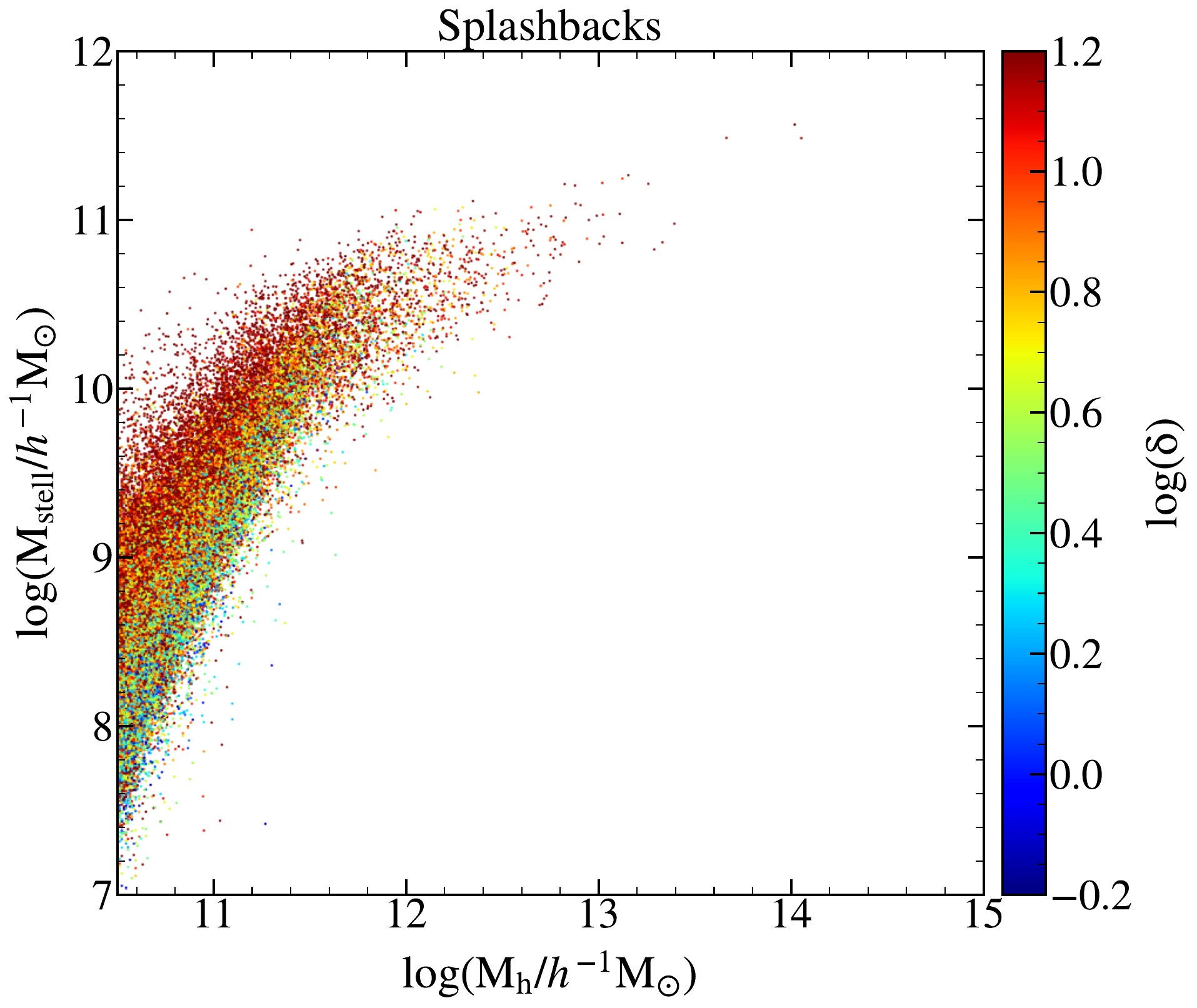}}
    \caption{The stellar-to-halo mass relations of semi-analytically modeled galaxies in the Millennium Simulation. The left column of subplots (a, c, e) is color-coded by halo concentration, and the right column of subplots (b, d, f) is color-coded by environment. The top row (a, b) shows the full population of central galaxies, the middle row (c, d) shows the centrals with splashbacks removed, and the bottom row (e, f) shows the relations for only the splashbacks. Compared to other centrals, splashbacks are seen to have low halo masses and relatively high stellar masses, and they occupy concentrated halos in dense environments.  For visual clarity, we include only $10\%$ of the total central galaxies, randomly chosen. 
    }
\label{Fig:SM_HM}
\end{figure*}

Figure~\ref{Fig:SM_HM} shows the stellar mass-halo mass relation of the central galaxies sample. The top row of panels shows the full sample, the middle row shows the sample without splashbacks, and the bottom row shows only the splashback galaxies. The left column of panels is color-coded by concentration, while the right column is color-coded by environment, with the density $\delta$ smoothed using a Gaussian filter on a scale of 1.25 $\hmpc$.

Overall, we see that the stellar mass increases with halo mass, with a significant scatter in the relation. The slope of the stellar mass-halo mass relation is affected by supernovae feedback at low halo mass and by AGN feedback at high halo mass \citep{Benson:2003a, Bower:2006}, with a resulting `knee' in the relation at a halo mass of around $10^{12}\hmsun$. Splashbacks are seen to reside largely below this knee, due to their relatively low halo mass. 

Examining the top two panels, we see the expected secondary trends with concentration and environment, consistent with previous work (e.g., \citealt{Zehavi:2018, Zehavi:2019}). Namely, at fixed halo mass, halos with higher concentration tend to host central galaxies with larger stellar masses. A similar trend exists also for environment at the low halo mass regime.
The splashback galaxies mostly populate low-mass halos that are highly concentrated, as can be seen in the bottom-left panel of Fig.~\ref{Fig:SM_HM}. This is likely a consequence of the arrested mass development and tidal stripping of the outer layers of the splashback halos, resulting in a concentrated core.
Splashbacks also appear to reside in relatively dense environments (bottom-right panel). The preference for splashbacks to occupy denser environments confirms the trend seen in Fig.~\ref{Fig:CW} and again follows from  the expected excess of halo interactions in dense regions. 

At fixed halo mass, splashbacks are seen to have higher stellar masses compared to other centrals. This is a further consequence of the evolution history of splashbacks which undergo mass stripping while passing through their former massive host halo. The mass loss is expected to be more dominant in the outer regions of the halo, and preferentially remove dark matter over stars (e.g., \citealt{Smith:2016}). This process naturally results with the splashbacks having higher stellar-to-halo mass ratios compared to other centrals. This trend might be impacted by the limitations of the semi-analytical model, as there is no explicit modeling of the stellar mass lost to tidal forces. Though the amount of stellar mass loss is expected to be minimal, this does potentially limit the accuracy of the stellar-to-halo mass estimate. However, our final results are robust against this uncertainty, given that we obtain similar findings for a hydrodynamic simulation (see Appendix~\ref{sec:appendix}). Our results are also consistent with \citet{Wetzel:2014} who used a $N$-body simulation with stellar mass assigned empirically to match the SDSS stellar mass function of \citet{Li:2009}. They find significantly larger stellar-to-halo mass ratios for the simulated ``ejected" satellites compared to all central galaxies.

The splashback galaxies seem to contribute significantly to the trends in the scatter of the stellar mass-halo mass relation, namely the dependence on secondary properties like halo concentration and environment. When removing the splashback galaxies, these dependences are less apparent at the low halo-mass end, as shown in the middle panels of Fig.~\ref{Fig:SM_HM}. This is strikingly so for the environment at the low-mass regime and partially so for concentration, where some dependence is still evident. The remaining concentration dependence may be related to similar tidal processes affecting galaxies near massive halos that have not crossed the virial radius. This is supported by the residual environment dependence.

With regard to the environment dependence, we note that there are two distinct regimes of the stellar mass-halo mass relation where the density is large (top-right panel of Fig.~\ref{Fig:SM_HM}). The low-mass end is largely explained by the splashback galaxies. However, even when removing the splashbacks (middle-right panel), we can see a clear dependence on environment at the high-mass end, with more massive galaxies/halos populating denser regions. This is simply a reflection of the well-studied dependence of the halo mass function on environment, with more massive halos residing in denser regions. This leads to a general picture where the primary environment dependence is due to halo mass and a secondary dependence due to the splashbacks. The latter again arises from the arrested mass development on the outskirts of neighboring halos in dense regions. It also supports the notion that splashbacks may be misclassified as central galaxies in low-mass halos instead of satellites of their former more massive host halos. 

Different hypotheses have been suggested to explain the physical origin of the concentration dependence of stellar mass at fixed halo mass.  Highly-concentrated halos tend to have formed earlier, allowing more time for accretion and star formation. Alternatively, more concentrated halos have deeper potential wells that can better retain the gas \citep{Matthee:2017, Wang:2026}. Our results here suggest that splashback galaxies and their associated physical processes may contribute to this trend.  Given the important role such secondary trends play in producing the occupancy variations and GAB (e.g., \citealt{Zehavi:2018}), we now proceed to investigate the impact of splashback galaxies in these phenomena.

\subsection{Occupancy Variations and Splashback Galaxies}

\begin{figure*}
 \centering
 \includegraphics[width=0.45\textwidth]{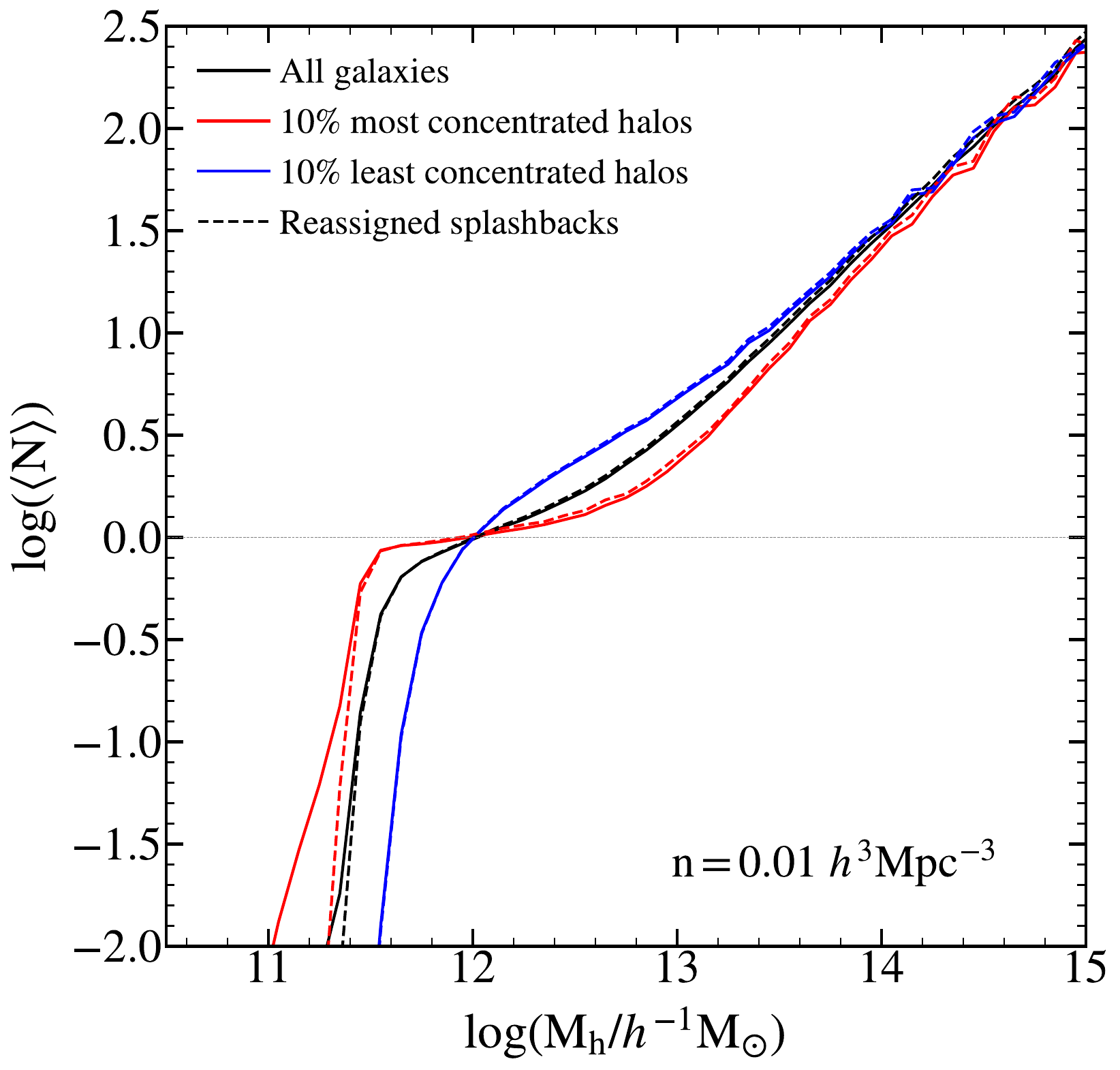} 
 \includegraphics[width=0.45\textwidth]{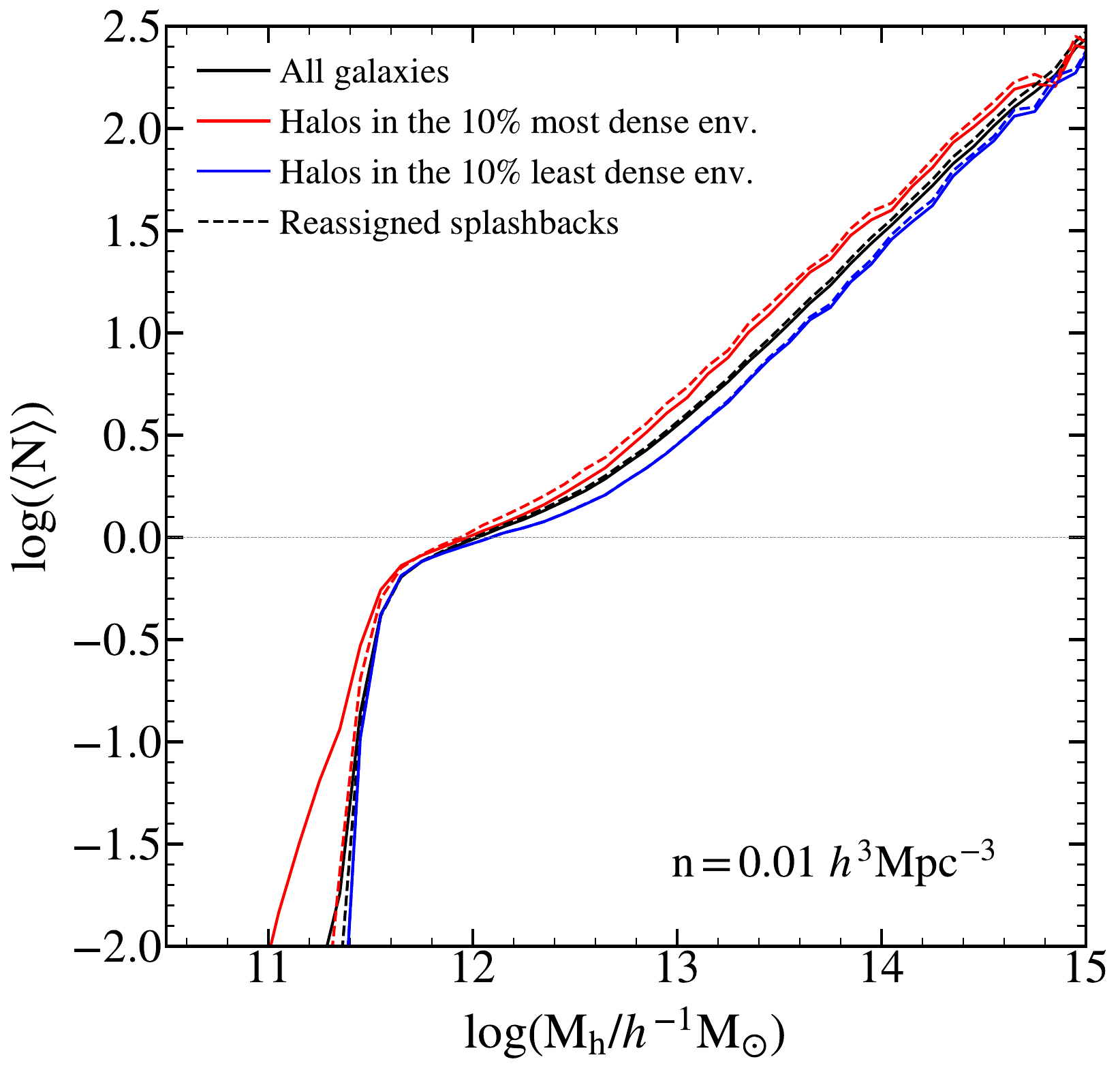}
 \caption{Halo occupation functions of semi-analytically modeled galaxies in the Millenium Simulation. The halo occupation function for the full galaxy sample corresponding to a number density of ${\rm n}=0.01 \hmpcc$ is shown as a solid black line in both panels. The left panel shows the occupancy variations with halo concentration, for the 10\% most concentrated halos (solid red line) and the 10\% least concentrated halos (solid blue line). The right panel shows the occupancy variations with environment, for the 10\% of halos in the densest environments (solid red line) and the 10\% of halos in the least dense environments (solid blue line). Dashed lines show the corresponding halo occupations after reassigning the splashbacks as satellites of their host halos (see text).} 
 \label{fig:hod}
 \end{figure*}

The halo occupation function, that is the mean number of galaxies as a function of halo mass, is a fundamental component of the HOD and characterizes the galaxy-halo connection of our galaxy samples. The halo occupation function may vary when restricting the halo population to a particular subset, such as the most or least concentrated halos, or halos in the densest or least dense environments. This dependence on secondary properties beyond mass, termed occupancy variation, may be indicative of galaxy assembly bias. Studying the impact of splashbacks on the occupancy variations of our galaxy samples provides insight into the possibility that splashback galaxies uniquely affect galaxy assembly bias.

The halo occupation functions for the stellar-mass-ranked galaxy sample corresponding to a number density of ${\rm n}=0.01 \hmpcc$ are shown in Figure \ref{fig:hod}. We first examine the overall occupancy variations before addressing the impact of splashbacks. The halo occupation function for the full sample is shown by the black solid line (in both panels). The left panel exhibits the occupancy variation with halo concentration, and the right panel shows the occupancy variation with environment.  The halo occupations are shown for the combined samples of both central and satellite galaxies. However, we can easily discern the central galaxies' occupation functions at the low-mass end and the satellites occupation function, which rise and dominate at high halo mass. 

Starting with environment, we see that galaxies in denser regions start populating lower-mass halos (red solid line in the right panel). This is evident by the central galaxies' mean occupation extending to lower halo masses and by a shift of the satellite mean occupation toward lower halo masses as well, reflecting preferential occupation of halos in dense environments, at fixed halo mass. 
We find significant trends with concentration as well. We see that more concentrated halos are more likely to host (central) galaxies at lower halo masses. The trend reverses around halo mass of about  $10^{12}\hmsun$, where the satellites occupation becomes dominant. At the high-mass end, we see that, at fixed halo mass, there are fewer galaxies in more concentrated halos. We can gain some intuition from the correlation between halo concentration and age. Concentrated early-formed halos have more time and stronger potential wells for developing central galaxies, while satellites have more time to merge with their centrals.  The reverse trends are seen for the least concentrated halos. 

These results, for both environment and concentration, are in agreement with the occupancy variations found in previous studies \citep{Artale:2018, Zehavi:2018, Bose:2019, Contreras:2019}.  The occupation variation for the central galaxies can also be qualitatively derived from the secondary trends in the stellar mass-halo mass relation. As demonstrated in Fig.~\ref{Fig:SM_HM}, at fixed halo mass, more concentrated halos and to some extent also halos in dense environments tend to host more massive galaxies. For fixed-number-density samples, this lead to occupancy variation where halos start hosting central galaxies at lower halo masses. See \citet{Zehavi:2018} for a more extensive discussion. 

We now turn to the impact of the splashback galaxies on the occupancy variations. The halo occupation functions for the galaxy samples with the reassigned splashbacks are plotted as dashed lines in Fig.~\ref{fig:hod}. The changes to the galaxy samples are two-fold. The first removes the splashbacks as central galaxies, thus altering the central galaxy occupation functions.  The second changes the satellite occupation function as the splashbacks are reassigned as satellites of their former host halos.  Hence, we can also easily consider here the case where the splashbacks are removed from the sample, by examining the changes to the central galaxy occupation at low mass, while keeping the satellite occupation at the high-mass end unchanged. 

The difference between the solid and dashed black lines (which are identical in the two panels of Fig.~\ref{fig:hod}) show the effect of the splashbacks reassignment for the full galaxy sample. Given the small fraction of splashbacks (see Table~\ref{tab:my_label}), the changes are very subtle but yet distinct. First, the reassignment removes central galaxies from the low-mass end of the halo occupation distribution. This finding is consistent with what we learn from Figs.~\ref{Fig:CW} and~\ref{Fig:SM_HM}, where splashbacks are shown to occupy predominantly low mass halos. This is also in agreement with \citet{Contreras:2013}, where splashbacks were found to dominate the tail of the central distribution. Secondly, the (satellites) halo occupation function slightly increases for halo masses larger than $\sim 10^{12}\hmsun$, as splashbacks are reassigned to higher mass
halos. 

For the occupancy variation with environment, shown in the right panel of Fig~\ref{fig:hod}, we see a significant change for the subpopulation of halos in the densest environments. The `tail' of galaxies in low-mass halos is mostly eliminated when the splashback galaxies are reassigned to their former host halos (as reflected in the change between the solid and dashed red lines). This reduces the level of occupancy variation to a large degree. In contrast, the low-mass end of the halo occupation in the least dense regions remains unchanged, with the dashed blue line overlapping the solid blue line. These behaviors are expected, as we have demonstrated in Fig.~\ref{Fig:SM_HM} that splashback galaxies predominantly occupy low-mass halos in dense environments, and that removing them eliminates most of the environmental dependence of the scatter at low halo mass. At high halo mass, we see a small increase in the occupation function after reassignment, mostly for halos in the densest environments, where the former hosts likely reside.

The effect of splashback reassignment on the occupancy variation with halo concentration, shown in the left panel of Fig.~\ref{fig:hod}, is similar to that of environment. At the low-mass end, where the occupation function mostly refers to central galaxies, the occupancy variation decreases for the most concentrated halos. The `tail' toward the very low-mass end is again essentially eliminated,  while the central galaxy occupation for the least concentrated halos remains unchanged. In contrast to the case of environment, significant differences in the central galaxy occupation functions remain for concentration.  Namely, the `turnover' location, or the minimum mass for hosting centrals in each of the (sub)samples, remains distinct even when removing the splashback galaxies. This is consistent with the remaining strong secondary trend with concentration seen in the stellar mass-halo mass relation for low halo masses, and the lack thereof for environment, as shown in the middle panels of Fig.~\ref{Fig:SM_HM}.
For higher halo masses, dominated by the satellites, the occupation functions slightly increase with the reassigned splashbacks, most noticeably for the most concentrated halos. 

We also examined the occupancy variations for our galaxy samples with number density of $0.00316 \hmpcc$ and $0.0316 \hmpcc$, finding very similar results. While the overall variations increase with number density, as expected (see, e.g., \citealt{Zehavi:2018}), the impact of the splashbacks and trends thereof are similar.

\subsection{The Effect of Splashback Galaxies on Galaxy Assembly Bias}
\label{sec:results.effect}

With the decreased occupancy variations when removing the splashback galaxies, it is particularly interesting to explore the impact on GAB.
For a given galaxy sample, we may assess  galaxy assembly bias from the population by randomly reassigning the galaxies among halos of the same mass, as explained in Section~3.2. This removes any dependence on secondary halo properties or the environment at large. The ratio of the correlation function of the galaxy sample to that of the shuffled sample measures the level of GAB. We study the GAB signal for our alternate treatments of splashback galaxies. 

\begin{figure*}
\centering
\includegraphics[width=0.45\textwidth, height=0.29\textheight]{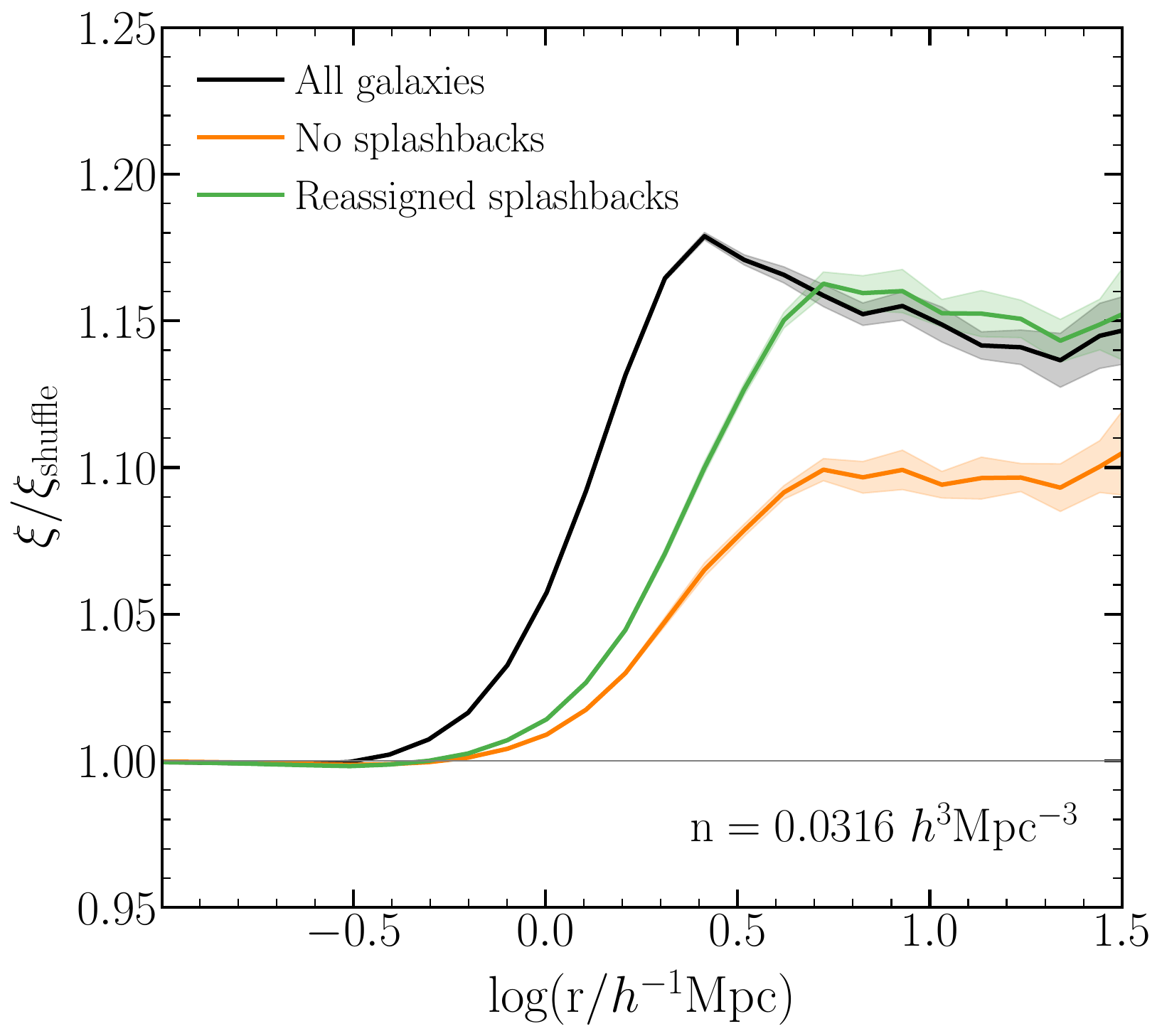}
\includegraphics[width=0.45\textwidth, height=0.29\textheight]{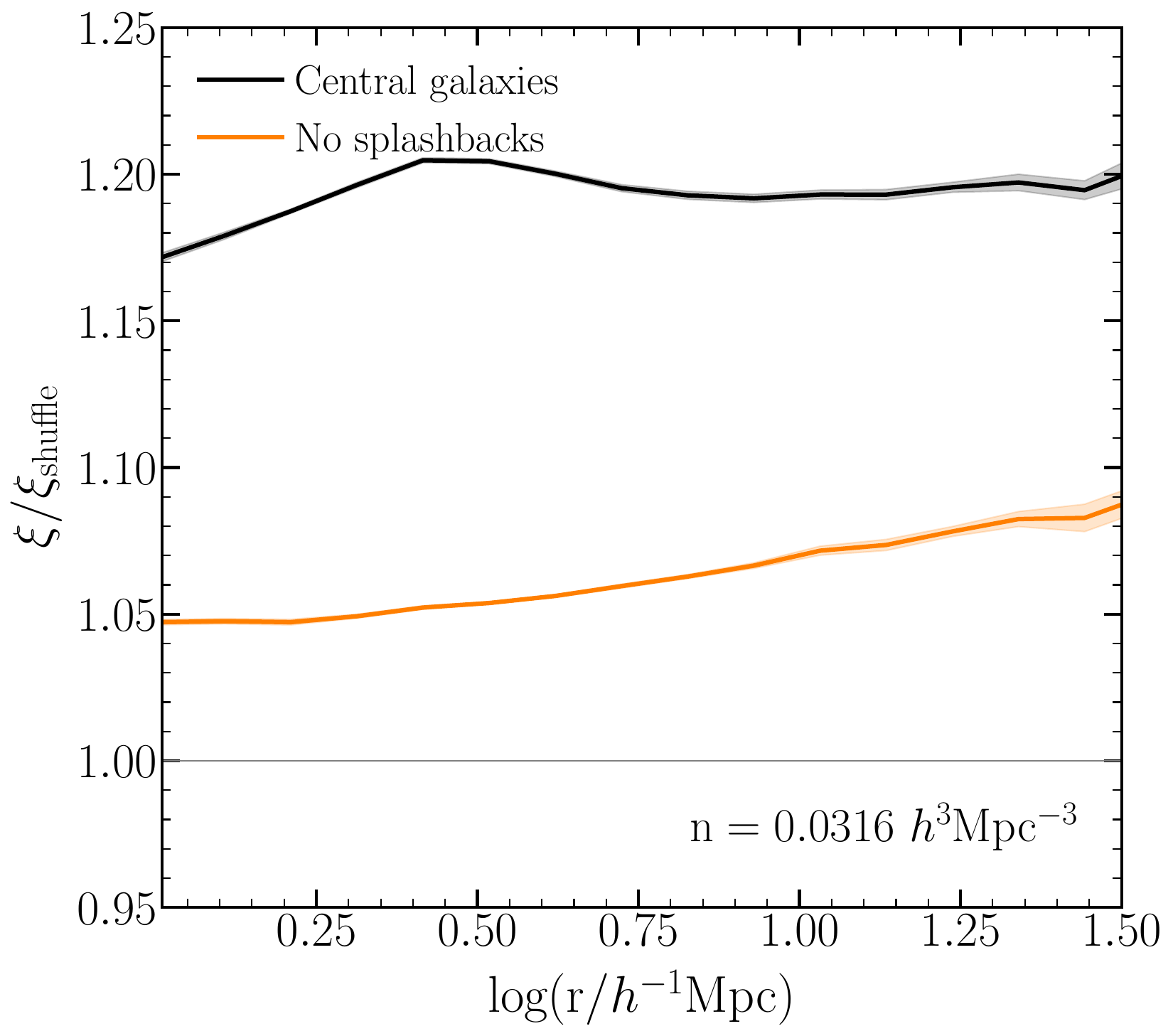}
\includegraphics[width=0.45\textwidth, height=0.29\textheight]{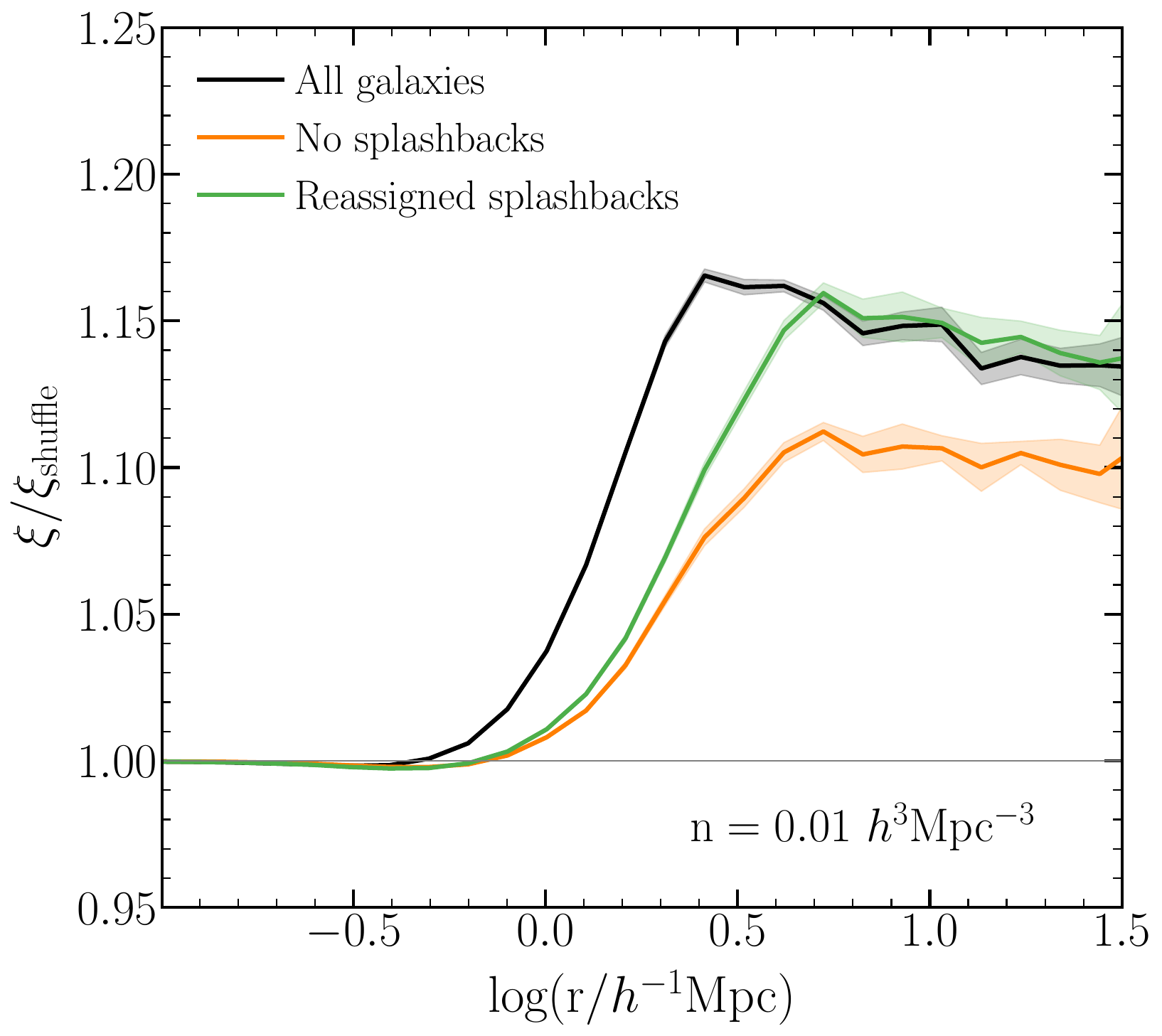}
\includegraphics[width=0.45\textwidth, height=0.29\textheight]{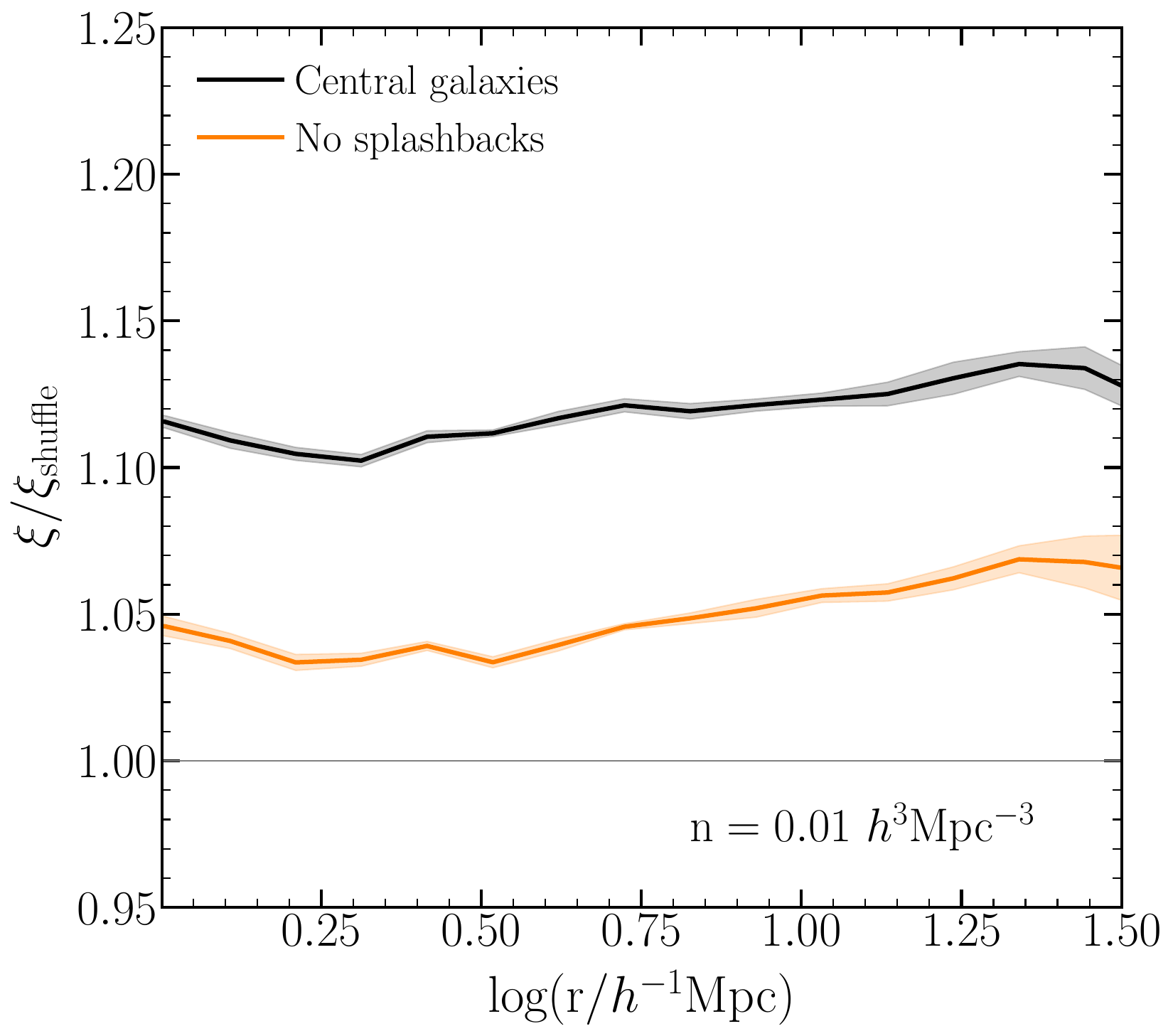}
\includegraphics[width=0.45\textwidth, height=0.29\textheight]{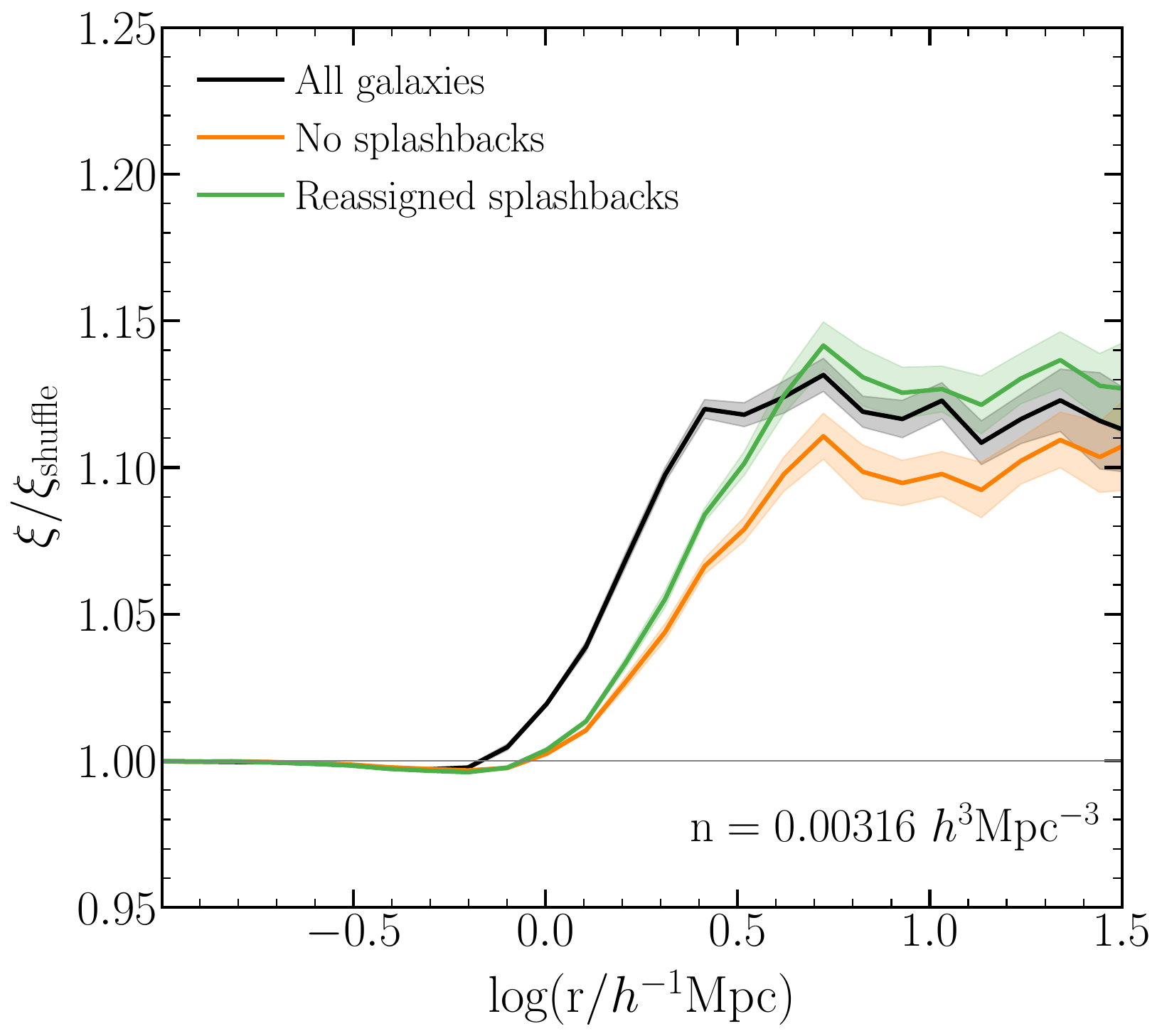}
\includegraphics[width=0.45\textwidth, height=0.29\textheight]{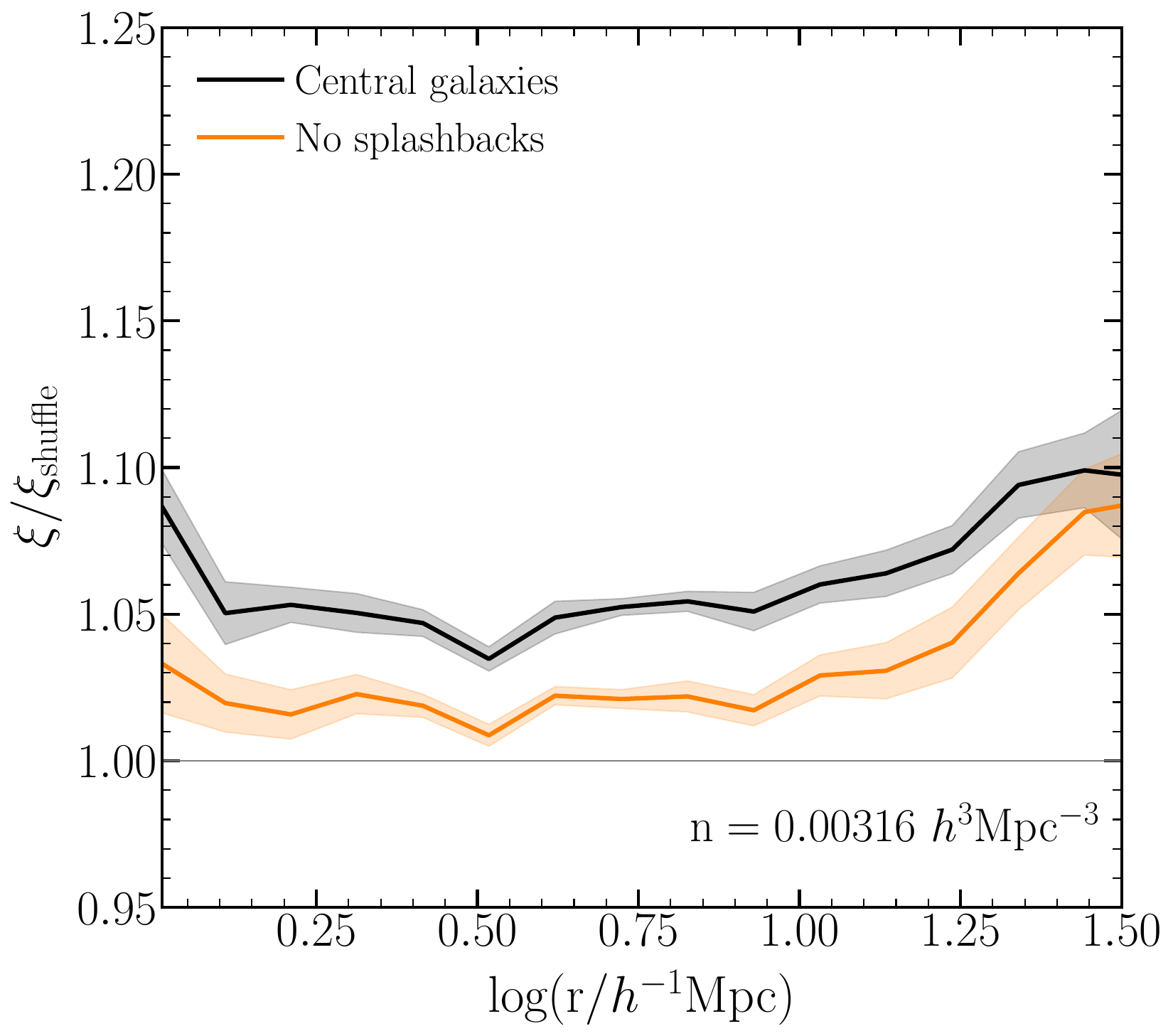}
\caption{Measured galaxy assembly bias for the ${\rm n}= 0.0316\hmpcc$ (top row), 0.01$\hmpcc$ (middle), and 0.00316$\hmpcc$ (bottom) galaxy samples. For each, the left panel shows correlation function ratios for the full (central and satellite) samples, while the right panel shows these ratios for centrals only (note that the latter has a different x-axis range). Solid lines correspond to the original samples (black), the samples with splashbacks removed (orange), and the samples with splashbacks reassigned as satellites (green). Shaded regions reflect the uncertainty from 10 different shuffling samples. We see that while removing the splashbacks makes a significant difference to the level of GAB, reassigning the splashbacks largely does not affect its amplitude.} 
\label{fig:shuffle}
\end{figure*}

Figure \ref{fig:shuffle} shows these correlation function ratios for the original galaxy sample (black), for the galaxy sample with the splashbacks removed (orange), and for the galaxy sample with the splasbacks reassigned (green). The shaded regions denote the uncertainty from the shuffling procedure, obtained from 10 shuffled samples.  The jackknife measurement uncertainties are expected to be subdominant (see, e.g., Figure~10 of \citealt{Zehavi:2018} or Figure~10 of \citealt{Contreras:2019}) and are therefore not shown. The three rows in Figure \ref{fig:shuffle} depict the correlation function ratios for our three different number density cuts, ${\rm n}=0.0316\hmpcc$ (top), $0.01\hmpcc$ (middle), and $0.00316\hmpcc$ (bottom). The two columns depict the ratios for the full galaxy population (left) and for centrals alone (right).

The correlation function ratios of the full samples (black lines) in the left panels of Fig.~\ref{fig:shuffle} show the expected GAB signature.  On small scales, the ratio is one by definition, as the intra-halo position of galaxies is not changed by the shuffling, leading to the same one-halo contribution to the correlation function. The correlation function ratio rises around $1 \hmpc$ and roughly plateaus at larger scales, exhibiting the effects of assembly bias. The large-scale correlation functions exhibit a $12-16\%$ increased amplitude relative to the shuffled samples, depending on the sample number density. The level of GAB increases slightly with the number density, or equivalently for lower stellar-mass thresholds. This trend is in agreement with previous work \citep{Croton:2007, Zehavi:2018, Contreras:2019, Contreras:2021a} that found stronger GAB for low-mass galaxies, corresponding to larger number-density samples. Additionally, the scale at which the ratio rises above unity shifts
to a lower value when increasing the number density, as the lower-mass galaxies included typically correspond to smaller halos.

When removing the splashback galaxies, we see a significant change in the amplitude of the correlation function ratios, shown as the orange lines in Fig.~\ref{fig:shuffle}.  The excess clustering relative to the shuffled samples is reduced to $\sim10\%$, with a larger relative reduction for the higher number density samples. While the majority of the GAB signal remains, it is still remarkable that removing about $3\%$ of the galaxies results in a roughly $30\%$ reduction of the large-scale signal for the ${\rm n}=0.01 \hmpcc$ sample. 
To focus on this effect, we compute the GAB signals when considering only the central galaxies in the samples, as is shown in the right column of Fig.~\ref{fig:shuffle}. In this case, removing the splashbacks reduces the GAB signal further, to a level of roughly $5\%$  excess clustering only, with larger relative reductions occurring again for the higher number-density samples. We find that, for the ${\rm n}=0.01$ sample, removing $\sim4\%$ of the central galaxies which are splashbacks, decreases the GAB signal by about $60\%$.

The significant reduction in the level of GAB when removing the splashback galaxies, particularly for the central galaxies, is consistent with the decreased occupancy variations discussed in \S4.2.  We note that the reduction of the GAB signal is not just due to the modified number density that comes with splashback removal. We have verified that increasing the stellar mass threshold of the original sample to match the number density of the removed sample does not produce a noticeable change in the correlation function ratio.
 
Given the unique assembly history of the splashback galaxies, we further examine the impact of changing their designation from central galaxies in their own halo to satellites of their former host halo. The impact of this reassignment is shown as the green lines in the left panels of Fig.~\ref{fig:shuffle}. For all three number density samples, we find that the level of assembly bias for the galaxies with the reassigned splashbacks does not differ significantly from that of the original sample. The GAB signal in these cases settles at a comparable, slightly larger ratio but is consistent within the uncertainty of the measurements.   The most noticeable change is that the scale of the transition from a ratio of unity to the increased GAB signature shifts to larger scales for the samples with the splashbacks reassigned, relative to that scale for the original samples. This is consistent with the splashbacks, which are at the outskirts of their former host halos, now being included in those halos, effectively increasing their size. 

We clarify that the reassignment of the splashback galaxies does not change the measured correlation function of the sample, since all galaxies retain their spatial positions, as noted in \S3.2. However, since their halo association changes, the resulting shuffled samples are different, albeit not enough to make a significant impact on the level of GAB.
The robustness of the GAB signal indicates that the potential misclassification of the splashback galaxies is not in itself the source of GAB. It also implies that studies of splashback-related assembly bias may be insensitive to how splashbacks are classified. In particular, different halo-finding algorithms may identify some halos as either splashbacks or as subhalos of larger neighboring halos. Our results suggest that such studies should be insensitive to the specific choice of halo finder.

To test the robustness of our results to the modeling of baryonic physics in the SAM, we perform a similar analysis of the impact of splashback galaxies in the TNG300 hydrodynamical simulation. Unlike in the case of a SAM, hydrodynamical simulations make no distinction of the prescription by which central galaxies and satellites evolve. Our results for the impact of TNG300 splashback galaxies on GAB, presented in Appendix \ref{sec:appendix}, are in agreement with those we find for the SAM, supporting the robustness of our conclusions.  

\section{Conclusion}
\label{sec:conclude}

In this work, we use a semi-analytical model applied to the Millennium Simulation to study the role of splashback galaxies in galaxy assembly bias. We analyze stellar-mass selected samples corresponding to three different number densities. Splashback galaxies are identified as current central galaxies that were formerly satellites in more massive halos.  We investigate their impact on assembly-bias related measures, when either removing the splashback galaxies or reassigning them to their former host halos. 

We find that splashback galaxies occupy the low mass end of the halo mass function and tend to reside in denser regions of the cosmic web, reflecting the arrested development of splashback halos in the vicinity of more massive halos. 
Examining the stellar mass-halo mass relation (Fig.~\ref{Fig:SM_HM}) we see that splashback galaxies have relatively high stellar-to-halo mass ratios. The splashback halos are highly concentrated and reside in dense environments, consistent with them contributing to assembly bias. These trends naturally follow as well from their truncated halo mass assembly. Interestingly, splashback galaxies appear to contribute significantly to the concentration and environment dependence of stellar mass at fixed halo mass. Our results therefore suggest that the physical processes leading to splashbacks may play an important role in these secondary dependences in the scatter of the stellar mass-halo mass relation.   

We study the impact of splashbacks on the occupancy variations with halo concentration and environment (Fig.~\ref{fig:hod}). Removing the splashback galaxies reduces the occupancy variation, mostly by eliminating the `tail' of the centrals occupation function toward low halo masses, exhibited for highly concentrated halos and separately for dense environments. Reassigning the splashbacks to their former host halos results in the same reduction in the central occupancy variation, but now with corresponding changes to the satellite occupation functions. The high-mass satellite occupancy variations, however, are not reduced by the splashback reassignment.

Most importantly, we compute the impact on the GAB signal, namely the ratio of the correlation functions of the original samples to those of shuffled samples. We examine it separately for the full samples, the samples where the splashback galaxies are removed,  and when reassigning the splashbacks, as shown in Fig.~\ref{fig:shuffle}.  Overall, we find that while removing the splashbacks makes a significant difference to the level of GAB, reassigning the splashbacks has little effect on its amplitude.
Splashbacks seem to account for a substantial fraction of GAB. When they are removed from the full galaxy sample, the signal is reduced by about a third for the ${\rm n}=0.01 \hmpc$ sample, with a stronger impact for higher number density (corresponding to a lower stellar-mass threshold). These results are qualitatively consistent with \citet{Mansfield:2020} who examined concentration-dependent halo clustering in a dark matter only $N$-body simulation and found that splashback halos account for a significant part of the halo assembly bias signal. A detailed comparison is not possible since halo assembly bias looks at differences in halo clustering for a specific halo property at fixed mass, while GAB measures the overall combined effect on galaxy clustering (from all halos and secondary properties). 

The key driver relating splashbacks to GAB appears to be the environmental effects that hinder the growth of the halos. We note that there may be no fundamental difference between true splashbacks that passed through another halo and galaxies that remained in the outskirts of massive halos. Such galaxies will undergo similar environmental effects and result in similar GAB. In that sense, splashbacks are not necessarily the origin of GAB, but rather tracers of it. 
One may be able to mitigate their effect by ``undoing" the arrested development along the lines of \citet{Lacerna:2011} and \citet{Smith:2024}. See also the related discussions in \citealt{Hahn:2009} and in Appendix B of \citealt{Contreras:2026}.  We reserve such investigations for future work.  

Our findings indicate that galaxy assembly bias is robust against the classification of splashback galaxies as centrals or as satellites of their prior host halos.
Different halo-finding algorithms may differ in their treatment of splashbacks,
but our results suggest that assembly bias studies may be insensitive to such choices (e.g., \citealt{Zehavi:2018}). 
To confirm that our overall conclusions are not dependent on the use of a semi-analytic model of galaxy formation, we repeat our analysis using the TNG300 hydrodynamical simulation, finding consistent results (see Appendix~\ref{sec:appendix}). 

Splashbacks remain a fascinating and important contributor to studies of assembly bias, with potential applications to modeling the galaxy-halo connection and ultimately for constraining cosmology.

\section*{Acknowledgements}

The Millennium simulation used in this work was carried out by the Virgo Consortium on the main supercomputer at the Max Planck Society's Supercomputing Centre in Garching, Germany. We gratefully acknowledge the German Astrophysical Virtual Observatory (GAVO) and the Virgo Consortium for making the simulation data available. We thank J\'onas Chaves-Montero, Shaun Cole and Kai Wang for useful discussions. I.Z. acknowledges support from a CWRU Expanding Horizons Initiative Finish Line Fund Award. S.C. acknowledges the support of the ``Ram\'on y Cajal'' fellowship (RYC2023-043783-I). S.C. also acknowledges the support of ``Ayudas para Atracci\'on de Investigadores con Alto Potencial'' (2025/00000640) from the Universidad de Sevilla.

\section*{Data Availability}

The Millennium Simulation databases used in this paper and the web application providing online access to them were constructed as part of the activities of the GAVO and are publicly available at \url{https://wwwmpa.mpa-garching.mpg.de/millennium/}
\citep{Springel:2005, Lemson:2006}. The IllustrisTNG simulations, including TNG300, are publicly available and accessible at \url{www.tng-project.org/data} \citep{Nelson:2019}. Data directly related to this publication are available upon reasonable request.

\appendix

\section{Splashback Galaxies in the TNG300 Hydrodynamical Simulation}
\label{sec:appendix}

To check the robustness of our results to the galaxy formation model used,   we repeat our analysis using the Illustris TNG300 hydrodynamical simulation \citep{Nelson:2019}.  Our results in the main part of the paper were all based on the \citet{Guo:2011} SAM applied to the Millennium Simulation.  The SAM follows specific physical prescriptions for galaxy formation and evolution, which are specific to whether the galaxy is a central galaxy or a satellite (for example, gas stripping in the circumgalactic medium once a galaxy and its corresponding (sub)halo become a satellite in a larger halo). In contrast, in hydrodynamical simulations the galaxy formation physics is agnostic to whether the galaxy is a central or a satellite. Hence, this test uses not only a completely different galaxy formation model but is also distinct in its treatment of central galaxies and satellites with respect to the SAM.

We utilize galaxy and halo samples from the TNG300 simulation, which is part of ``The Next Generation'' Illustris suite of hydrodynamical cosmological simulations of galaxy formation \citep{Pillepich:2018a, Pillepich:2018b, Springel:2018, Nelson:2019}, a successor of the original Illustris simulation \citep{Vogelsberger:2014a, Vogelsberger:2014, Genel:2014}.
We use their largest high-resolution simulation, TNG300, which has a side length of $L = 205 h^{ -1}\text{Mpc}$, with periodic boundary conditions. The simulation tracks the evolution of $2500^3$ gas cells and $2500^3$ dark matter particles from redshift $z = 127$ to $z = 0$, with a baryon mass resolution of $7.4 \times 10^6 h^{-1}M_{\odot} $ and a dark matter particle mass of $4.0 \times 10^7 h^{-1}M_{\odot}$. The identification of halos and (central and satellite) galaxies and their stellar mass content is detailed in \citet{Pillepich:2018b}. The outputs are publicly available on the TNG project website\footnote{\url{https://www.tng-project.org/}}.

We apply a stellar-mass threshold of $0.83 \times 10^{10} h^{-1}M_{\odot}$
to galaxies in the TNG300 simulation, corresponding to ${\rm n}=0.01\hmpcc$. We identify splashback galaxies using the same definition applied to the Millennium Simulation galaxy samples, and detailed in Section \ref{sec:methods.definition}. The TNG300 ${\rm n}=0.01\hmpcc$ galaxy sample includes 2179 splashback galaxies, which amount to 2.1\% of all galaxies and 3.4\% of the central galaxies.  We note that, given the large difference in simulation volume, the TNG300 galaxy sample is much smaller than the corresponding SAM sample (which includes $\sim 33,000$ splashback galaxies), precluding a detailed comparison. Nonetheless, we proceed with a qualitative comparison of the GAB signal. 

\begin{figure}
\centering
\includegraphics[width=0.45\textwidth, height=0.29\textheight]{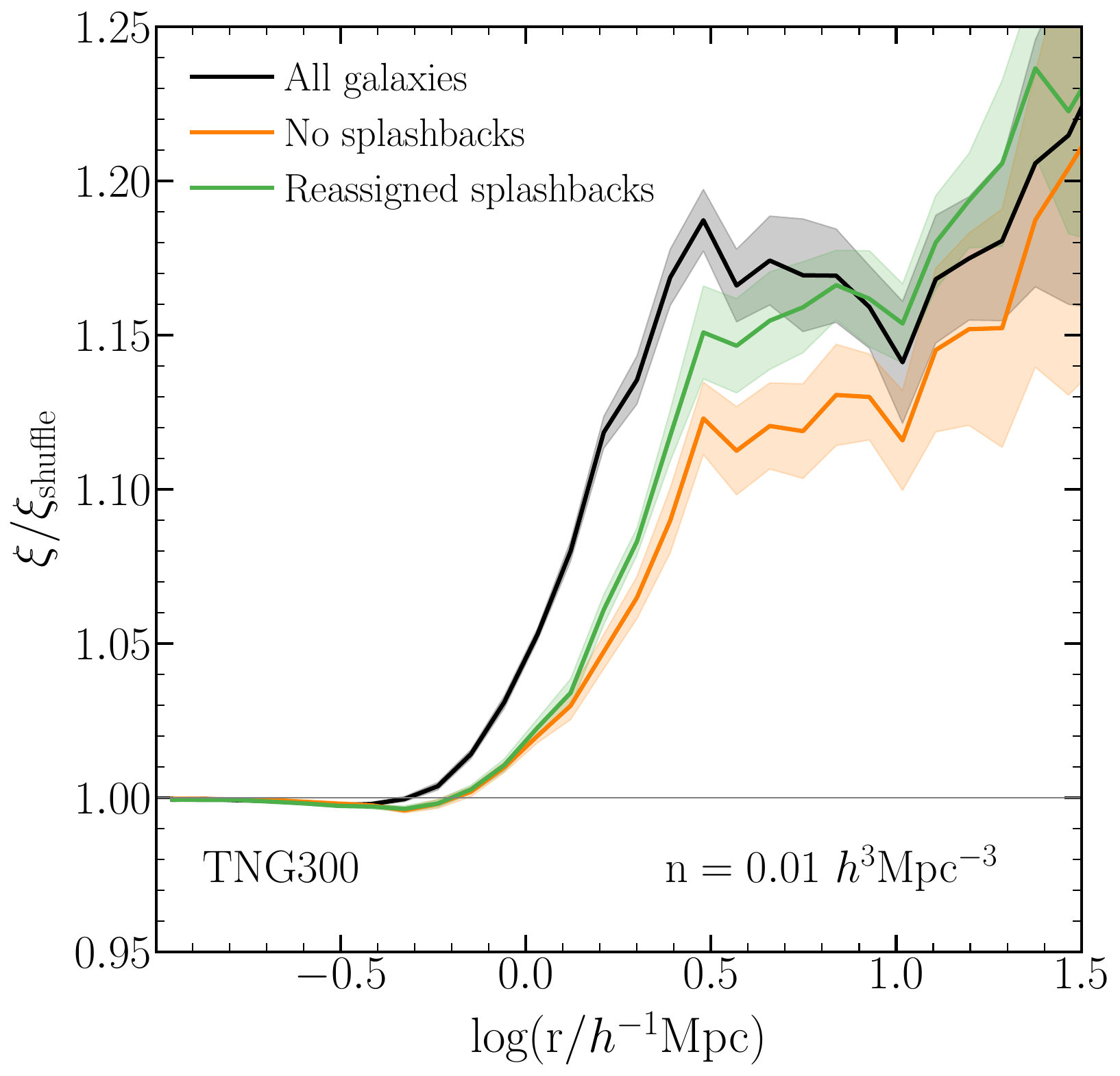}
\includegraphics[width=0.45\textwidth, height=0.29\textheight]{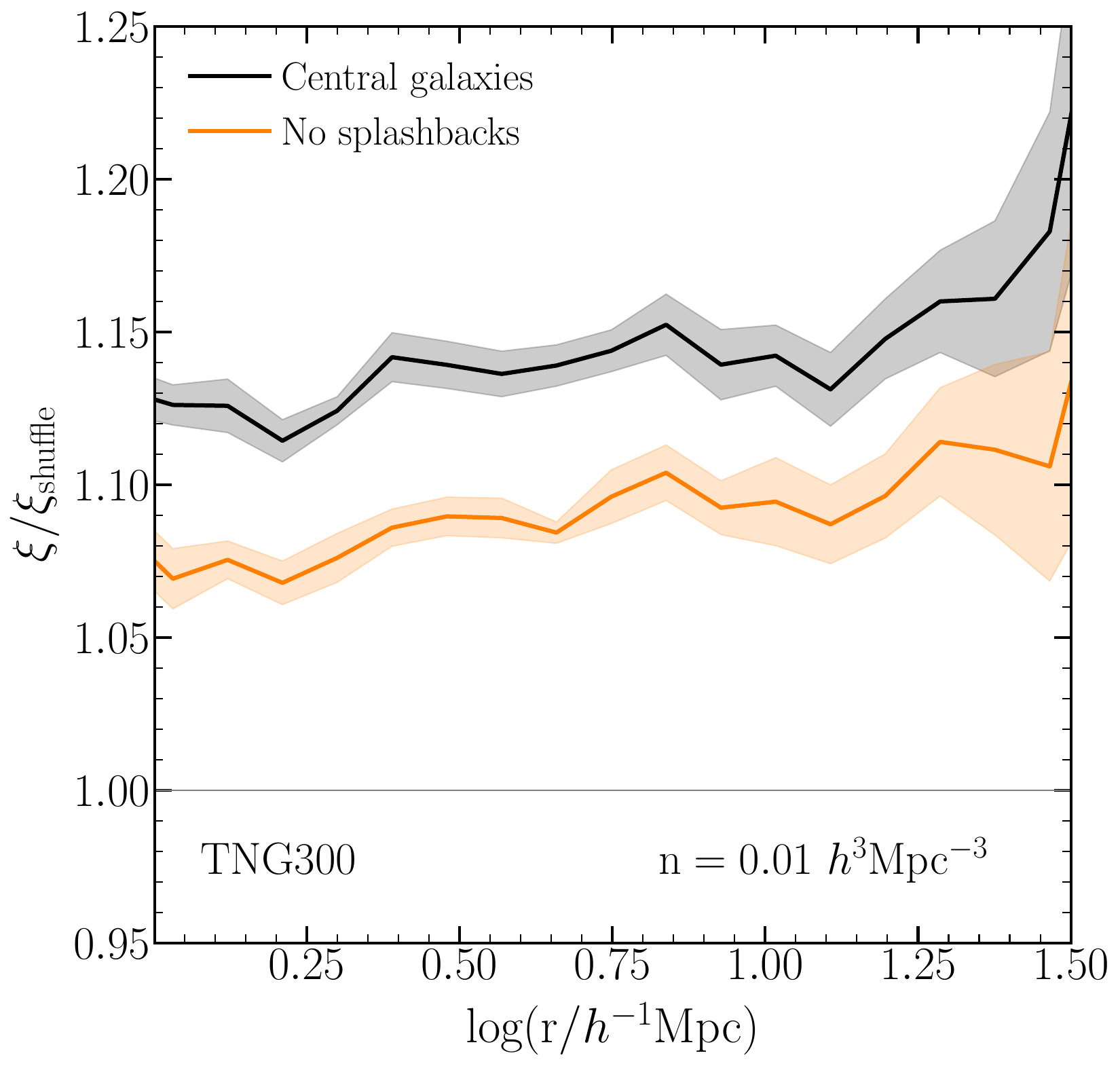}
\caption{Galaxy assembly bias measurements for the ${\rm n} = 0.01\hmpcc$ TNG300 galaxy sample. Correlation function ratios for the full (central and satellite) samples are shown on the left, while the right panel shows these ratios for the central galaxies only.  Solid lines represent the original samples (black), the samples with the splashbacks removed (orange), and the samples with splashbacks reassigned to their former hosts (green). The shaded regions denote the uncertainty from 10 different shuffling samples. The results are consistent with our conclusions from the SAM, within the larger uncertainties for the TNG300.}
\label{fig:tng}
\end{figure}

We assess the strength of the GAB signal using the shuffling test when splashback galaxies are either removed or reassigned as satellites of their former host halos.
Figure~\ref{fig:tng} shows the assembly bias signal of the TNG300 for all galaxies in the top panel and for central galaxies only in the bottom panel. As before, the black line is the ratio of the clustering of the original full sample to that of the shuffled sample (without assembly bias). The orange line shows the clustering ratio when the splashback galaxies are removed and the green line corresponds to the galaxy sample with the splashback galaxies reassigned to their former halos. The shaded regions represent the uncertainty obtained from 20 different randomly shuffled samples. 

As with the SAM, we find that the GAB signal is noticeably smaller without the splashback galaxies. The signal is reduced by about $30\%$ for all galaxies and by roughly $40\%$ for only the centrals. However, most of the GAB signal remains intact here as well. When reassigning the splashback galaxies, the signal remains roughly the same as that of the original sample (within the error bars), again demonstrating that GAB is insensitive to the treatment of the splashback galaxies -- whether they are regarded as central galaxies or satellites of their former host halos. We found results similar to those from the SAM also for the occupancy variations and for the stellar mass-halo mass relation in TNG300 (not shown). Overall, while noisier than the SAM results, the TNG300 results are consistent with them, demonstrating an important robustness to the galaxy formation model. 

\bibliography{Biblio}{}
\bibliographystyle{aasjournal}

\end{document}